\documentclass[11pt]{article}

\usepackage[margin=.75in]{geometry}

\usepackage{amsmath,amssymb}
\usepackage{cancel}
\usepackage{graphicx}
\usepackage{caption}
\usepackage{comment}
\usepackage{subcaption}

\usepackage{nicematrix}
\usepackage{bm}

\usepackage{color}

\definecolor{indigo}{RGB}{0,0,120}
\usepackage[colorlinks=true, linkcolor=indigo, citecolor=blue, urlcolor=indigo,backref=page]{hyperref}

\def\tr{\,{\rm tr}\,}

\def\fl{\noindent}

\newcommand{\bra}{\langle}
\newcommand{\ket}{\rangle}

\newcommand{\tl}[1]{\tilde{#1}}
\newcommand{\dd}[2]{\frac {\partial #1}{\partial #2}}

\newcommand{\pdr}{\partial}

\newcommand{\beq}{\begin{equation}}
\newcommand{\eeq}{\end{equation}}
\newcommand{\beqs}{\begin{eqnarray}}
\newcommand{\eeqs}{\end{eqnarray}}
\newcommand{\half}{\frac{1}{2}}
\newcommand{\ov}[1]{\frac{1}{#1}}

\def\al{\alpha}

\def\g{\gamma} 
\def\eps{\epsilon} 
\def\la{\lambda}
		
\def\sig{\sigma}
\def\tht{\theta}
\def\om{\omega}

\newcount\colveccount  % column vec \colvec{nrows}{a11 & a12}{a21 & a22}
\newcommand*\colvec[1]{\global\colveccount#1  \begin{pmatrix} \colvecnext} \def\colvecnext#1{#1 \global\advance\colveccount-1
        \ifnum\colveccount>0 \\ \expandafter\colvecnext
        \else \end{pmatrix} \fi}

\usepackage{calligra} % for script and cursive characters like script r
\DeclareMathAlphabet{\mathcalligra}{T1}{calligra}{m}{n}
\DeclareFontShape{T1}{calligra}{m}{n}{<->s*[2.2]callig15}{}

\begin{document}
%---------------------------

\title{
\hfill{\tt \small \href{https://arxiv.org/abs/2111.03858}{arXiv:2111.03858 [math-ph]}}\\
Quantum Rajeev-Ranken model as an anharmonic oscillator}
\author{{\sc Govind S. Krishnaswami and T. R. Vishnu}
\\ \small
Physics Department, Chennai Mathematical Institute,  SIPCOT IT Park, Siruseri 603103, India
\\ \small
Email: {\tt govind@cmi.ac.in, vishnu@cmi.ac.in}}
\date{November 19, 2021\\
Published in \href{https://doi.org/10.1063/5.0079269}{J. Math. Phys. 63, 032101 (2022)}}

\maketitle

\abstract{ \small The Rajeev-Ranken (RR) model is a Hamiltonian system describing screw-type nonlinear waves of wavenumber $k$ in a scalar field theory `pseudodual' to the 1+1D SU(2) principal chiral model. Classically, the RR model is Liouville integrable. Here, we interpret the model as a novel 3D cylindrically symmetric quartic oscillator with an additional rotational energy. The quantum theory has two dimensionless parameters. Upon separating variables in the Schr\"odinger equation, we find that the radial equation has a four-term recurrence relation. It is of type $[0,1,1_6]$ and lies beyond the ellipsoidal Lam\'e and Heun equations in Ince's classification. At strong coupling $\la$, the energies of highly excited states are shown to depend on the scaling variable $\la k$. The energy spectrum at weak coupling and its dependence on $k$ in a double-scaling strong coupling limit are obtained. The semi-classical Wentzel-Kramers-Brillouin (WKB) quantization condition is expressed in terms of elliptic integrals. Numerical inversion enables us to establish a $(\la k)^{2/3}$ dispersion relation for highly energetic quantized `screwons' at moderate and strong coupling. We also suggest a mapping between our radial equation and one of Zinn-Justin and Jentschura that could facilitate a resurgent WKB expansion for energy levels. In another direction, we show that the equations of motion of the RR model can also be viewed as Euler equations for a step-3 nilpotent Lie algebra. We use our canonical quantization to uncover an infinite dimensional reducible unitary representation of this nilpotent algebra, which is then decomposed using its Casimir operators.}

%---------------------
\vspace{0.25 cm}
%---------------------
\footnotesize

{\fl \bf Keywords:} anharmonic oscillator, separation of variables, dispersion relation, Euler equations, WKB approximation, nilpotent Lie algebra.

\normalsize

{\scriptsize \tableofcontents}

\normalsize

%------------------------------------
\section{Introduction}
\label{s:Introduction}
%------------------------------------

In this paper, we study some aspects of the quantum Rajeev-Ranken (RR) model. The RR model is a mechanical system with three degrees of freedom describing a class of nonlinear waves in a 1+1-dimensional (1+1D) scalar field theory that was introduced in the work of Zakharov and Mikhailov \cite{Z-M} and Nappi \cite{Nappi}. It is `pseudodual' to the 1+1D SU(2) principal chiral model (PCM), which is equivalent to the 1+1D SO(4) nonlinear sigma model. The latter is an effective theory for pions, displays asymptotic freedom, possesses a mass gap \cite{Polyakov} and serves as a toy-model for 3+1D Yang-Mills theory. The PCM and nonlinear sigma model are examples of integrable field theories and nonperturbative results concerning their $S$-matrix and spectrum have been obtained using the methods of integrable systems \cite{Z-Z, P-W, F-R}.

Unlike the PCM, which is based on a Euclidean current algebra, its pseudodual scalar field theory (which is obtained via a noncanonical change of variables) is based on a step-3 nilpotent current algebra  (all thrice-iterated brackets vanish). Moreover, the quantum theories are quite different. In fact, the scalar field theory is strongly coupled in the ultraviolet. Thus, as pointed out by Rajeev and Ranken {\cite{R-R}}, it could serve as a lower-dimensional toy-model for studying certain nonperturbative aspects of theories with a perturbative Landau pole (such as 3+1D $\la \phi^4$ theory). In particular, one wishes to identify degrees of freedom appropriate to the description of the dynamics of such models at high energies (if indeed a UV completion can be defined). Although the nilpotent scalar field theory has been shown by  Curtright and Zachos \cite{C-Z} to possess infinitely many {\it nonlocal} conservation laws, it has not yet been possible to solve it in anywhere near the way that the PCM has been solved. However, the scalar field theory shares with the harmonic and anharmonic oscillators, as well as field theories such as Maxwell, $\la \phi^4$ and Yang-Mills, the feature of being based on a nilpotent Lie algebra and a quadratic Hamiltonian.  Thus, it is plausible that some common approximation methods and techniques of solution may apply to several of these models.

As a step towards understanding this scalar field theory, Rajeev and Ranken \cite{R-R} reduced it to a Hamiltonian system with three degrees of freedom describing a class of nonlinear constant energy-density classical waves. These novel `screw-type' continuous waves, which we call {\it screwons}, could play a role similar to solitary waves in other field theories and it is important to understand their quantum nature. In this paper, we address this question and some other aspects of the quantum RR model by interpreting it as a quartic oscillator. A brief survey of some of the literature on quantum anharmonic oscillators may be found in \S \ref{s:Quantum-RR-model}.

This work builds on our earlier investigations into the classical integrability and dynamics of the RR model. The latter bears some resemblance to the Neumann (for a particle moving on a 2-sphere) and Kirchhoff models, but is different from them. In \cite{G-V-1, G-V-2}, we found a Lagrangian and a pair of Hamiltonian formulations for the RR model, based on compatible degenerate nilpotent and Euclidean Poisson algebras. Lax pairs and classical $r$-matrices were found. Casimir invariants were used to identify all symplectic leaves. On these leaves, a complete set of independent conserved quantities in involution was found, establishing the Liouville integrability of the model. The phase space is foliated by invariant tori, which arise as common level sets of conserved quantities. The latter are generically 2-tori, though horn tori, circles and points also arise, depending on the nature of roots of a cubic polynomial. On the nongeneric tori, conserved quantities develop relations and solutions degenerate from elliptic to hyperbolic, circular and constant functions. We also discovered a family of action-angle variables that are valid away from horn tori (where the dynamics was expressed as a gradient flow).

In \S \ref{s:Electromagnetic-interpretation-of-the-RR-model}, we begin with Rajeev and Ranken's interpretation of their model in terms of a charged particle moving in a static electromagnetic field \cite{R-R}. They quantized the model and applied dimensional arguments to their radial Schr\"odinger equation to deduce dispersion relations for the quantized screwons in weak and strong coupling limits. However, their radial equation appears to have an error and we have not been able to reproduce their dimensional argument in the strong coupling limit. In \S \ref{s:Quantization-of-the-electromagnetic-Hamiltonian}, we derive what we believe are the corrected equations of this electromagnetic interpretation. In \S \ref{s:Mechanical-interpretation-of-the-Rajeev-Ranken-Model}, we take a complementary approach by interpreting the RR model as a novel 3D cylindrically symmetric anharmonic oscillator. This interpretation follows from rewriting the Hamiltonian in terms of the Darboux coordinates introduced in \cite{G-V-1, G-V-2} and identifying the coordinates and momenta as those of a nonrelativistic particle subject to a quartic potential and possessing an unusual rotational energy. The parameters of the RR model (the coupling constant $\la$ and screwon wavenumber $k$) enter the oscillator in an intricate way, which leads to an unconventional physical interpretation of solutions of the oscillator. In \S \ref{s:Quantum-RR-model}, we exploit this mechanical interpretation to canonically quantize the model and observe that the quantum theory involves two dimensionless parameters constructed from $\la$ and $\hbar$. It is exactly solvable to the extent that one may completely separate variables in the  Schr\"odinger equation. The resulting radial equation (of type $[0,1,1_6]$) is a confluent form of an equation with 10 elementary regular singular points  and lies beyond the ellipsoidal Lam\'e  and Heun equations, which involve 5 and 8 elementary regular singularities (see Appendix \ref{A:Singularities-of-second-order-ordinary-differential-equations}). On the other hand, the 2-sphere quantum Neumann model \cite{B-T, Be-T} leads to an ordinary differential equation (ODE) of type $[3,0,1_1]$, arising from 6 elementary regular singularities.

We obtain the energy spectrum at weak coupling and its dependence on the wavenumber in a double-scaling strong coupling limit. In a more conventional strong coupling limit ($\la \to \infty$), we show that the energies of highly excited states can depend on $\la$ and $k$ only through their product. In \S \ref{s:Semi-classical-WKB-approximation}, we  find the WKB quantization condition in an implicit form. Although it leads to the expected weak coupling spectrum, we have not yet been able to obtain the spectrum explicitly for general values of the coupling. However, we determine the WKB spectrum numerically and find the dispersion relation $E_n \propto (\la k)^{2/3}$ for highly excited screwons at moderate to strong coupling. In \S \ref{s:Discussion}, we comment on the possibility of an `exact' WKB analysis, by relating our radial equation and the parameters of the RR model to those that appear in the work of Zinn-Justin and Jentschura \cite{Z-J-1, Z-J-2}. In another direction, we notice that the equations of motion (EOM) of the RR model can also be interpreted as Euler equations for a step-3 nilpotent Lie algebra. In \S \ref{s:unitary-representation-of-nilpotent-Lie-algebra}, we exploit our canonical quantization to uncover an infinite dimensional reducible unitary representation of this nilpotent algebra, which is then decomposed using its Casimir operators. \S \ref{s:Discussion} contains a discussion of our results and some open questions. Some of the results in this paper have appeared in PhD thesis \cite{Vishnu}.

%----------------------------------------------
\section{Electromagnetic interpretation of the RR model}
\label{s:Electromagnetic-interpretation-of-the-RR-model}
%-----------------------------------------------

Before viewing the RR model as an anharmonic oscillator, we revisit the interpretation given by Rajeev and Ranken \cite{R-R} in terms of a charged particle moving in an electromagnetic field. In the process, we find and rectify an unfortunate error in their equations. 

The RR model concerns the dynamics of screw-type waves in a scalar field theory for the $\mathfrak{su}(2)$ Lie algebra valued field $\phi(x,t)$ with the EOM $\ddot \phi - \phi'' = \la [\dot \phi, \phi']$, where  $\la > 0$ is a dimensionless coupling. These waves arise via the ansatz $\phi(x,t) = e^{Kx} R(t) e^{-Kx} + m K x$. Here, $m$ is a dimensionless parameter and $K = i k \sig_3/2$, where $k$ is a real parameter with dimensions of a wavenumber. The degrees of freedom in the  $\mathfrak{su}(2)$ matrix $R(t)$ can be taken as $R_a = i \tr (R \sig_a)$, where $\sig_a$ for $a= 1,2,3$  are the Pauli matrices. The Hamiltonian of the RR model in terms of $R_a$ and their conjugate momenta (see Eqn.~(41) of \cite{G-V-1}): 
	\beq
	kP_{1,2} = \dot R_{1,2} \pm \half \la k m R_{2,1} \quad \text{and} \quad kP_3 = \dot R_3 + \half \la k (R_1^2 + R_2^2)
	\label{e:conjugate-momenta}
	\eeq 
is:
	\beq
	 \frac{H}{k^2} = \sum_{a=1}^3 \frac{P_a^2}{2} + \frac{\la m}{2} \left( R_1 P_2 - R_2 P_1 \right) + \frac{\la^2}{8} \left( R_1^2 + R_2^2 \right) \left[ R_1^2 + R_2^2 + m^2 - \frac{4}{\la} \left( P_3 - \frac{1}{\la} \right) \right] + \frac{m^2}{2}.
	 %H &=& \half \left[ \left( k P_1 - \frac{\la m k R_2}{2} \right)^2 + \left( k P_2 + \frac{\la m k R_1}{2} \right)^2 + \left( k P_3 - \frac{\la k}{2}(R_1^2 + R_2^2) \right)^2 \right] + \frac{k^2}{2}(R_1^2 + R_2^2 + m^2). \qquad
	\label{e:Hamiltonian-mech-Darboux}
	\eeq

%---------------------------------------------	
\subsection{Classical charged particle in an axisymmetric field}
\label{s:Classical-charged-particle-in-an-axisymmetric-field}
%---------------------------------------------
	
Suppose that the Cartesian position and momentum coordinates of a charged particle are $x, y, z = R_{1,2,3}$ and $p_{x,y,z} = k P_{1,2,3}$, then the Hamiltonian in (\ref{e:Hamiltonian-mech-Darboux}) can be rewritten as:
	\beq
	H = \frac{1}{2 \mu} \left[ \left( p_x - \frac{q \la k m y}{2c} \right)^2 + \left( p_y + \frac{q \la k m x}{2c} \right)^2 + \left( p_z - \frac{q \la k}{2c}(x^2 + y^2) \right)^2 \right] + \frac{q k^2}{2}(x^2 + y^2 + m^2),
	\label{e:Hamiltonian-EM-Cartesian}
	\eeq
in units where the particle has mass $\mu = 1$, charge $q =1$ and the speed of light $c =1$. This describes a charged particle moving in a static EM field with the vector and scalar potentials: 
	\beq
	{\bm A} = \frac{\la k}{2} \left( m y, -m x, x^2 + y^2 \right) \quad \text{and} \quad 
	V(x, y, z) = \frac{k^2}{2}(x^2 + y^2 + m^2).
	\eeq
The ${\bm E}$ field points radially inwards while  ${\bm B}$ has both azimuthal and axial components:
	\beq
	\bm E = -k^2 ( x \hat x + y \hat y) \quad \text{and} \quad 
	\bm B = \la k \left(  y \hat x - x \hat y - m \hat z \right).
	\eeq
The Hamiltonian in (\ref{e:Hamiltonian-EM-Cartesian}) along with the canonical PBs $ \{ x, p_x \} = 1$ etc., gives the Newton-Lorentz (NL) equations  $\mu \ddot {\bm r} = q ({\bm E} + ({\bm v} \times {\bm B})/c)$. In fact, using 
	\beq
	\dot x = \ov{\mu} \left( p_x - \frac{q \la k m y}{2 c} \right), \quad \dot y = \ov{\mu} \left( p_y + \frac{q \la k m x}{2 c} \right) \quad \text{and} \quad \dot z = \ov{\mu} \left( p_z - \frac{q \la k}{2 c}(x^2 + y^2) \right), 
	\eeq
we get the NL equations in component form
	\beqs
	\mu \ddot{x} &=& -q k^2 x + \frac{q}{c} \la k \left( -m \dot{y} + x \dot{z} \right), \quad
	\mu \ddot{y} = -q k^2 y + \frac{q}{c} \la k \left( m \dot{x} + y \dot{z} \right)  \quad \text{and} \cr
	\mu \ddot{z} &=& -\frac{q}{c} \la k \left( x \dot{x} + y \dot{y} \right).  \qquad
	\label{e:EOM-EM-component-form}
	\eeqs
These agree with the EOM of the RR model following from (\ref{e:Hamiltonian-mech-Darboux}) in units where $\mu = q= c = 1$.

%---------------------------------------------
\vspace{0.25cm}
%---------------------------------------------

{\bf \fl Classical Hamiltonian in terms of cylindrical coordinates:} The Hamiltonian (\ref{e:Hamiltonian-EM-Cartesian}) is invariant under rotations about and translations along the $z$-axis. Hence, we make a canonical transformation to cylindrical coordinates $r = \sqrt{x^2 + y^2}, \tht = \arctan(y/x)$, and $z$ and their conjugate momenta $p_r = (x p_x + y p_y)/r $, $p_\tht = -y p_x + x p_y$ and $p_z$ satisfying $\{ r, p_r\} =1$ etc. The Hamiltonian corresponding to  (\ref{e:Hamiltonian-EM-Cartesian}) is
	\beq
	H = \frac{1}{2 \mu} \left[ p_r^2 +  \frac{1}{r^2}\left( p_{\tht} - \frac{q A_\tht}{c} \right)^2 + \left(p_z - \frac{q A_z}{c} \right)^2 \right] + q V(r),
	\label{e:Hamiltonian-EM-plane-polar}
	\eeq
where the vector and scalar potentials are
	\beqs
	{\bm A} = \frac{\la k}{2} (-m r \hat{\tht} + r^2 \hat{z})  & \text{and} &  V(r) = \frac{k^2 (r^2 + m^2)}{2}  \cr
	\text{with} \qquad A_z = {\bm A} \cdot \hat z = \frac{\la k r^2}{2}
	& \text{and} &
	A_\tht = r {\bm A} \cdot \hat \tht = -\frac{\la k m  r^2}{2}.
	\label{e:canonical-components-of-electro-magnetic-potentials}
	\eeqs
(In \cite{R-R}, while computing the effective potential, $A_\tht$ was mistakenly taken as $\la k m r/2$.) The resulting electric and magnetic fields are 
	\beq
	{\bm E} = -k^2 r \hat{r} \quad \text{and} \quad {\bm B} = -\la k \left( r \hat{\tht} + m \hat{z} \right).
	\eeq
In terms of velocities, the conjugate momenta are 
	\beq
	p_{r} = \mu \dot r, \qquad
	p_{\tht} = \mu r^2 \dot \tht + \frac{q A_{\tht}}{c} \qquad \text{and} \qquad
	p_z = \mu \dot z + \frac{q A_z}{c}.
	\eeq
Note that $p_{\tht} = r {\bm p} \cdot \hat \tht$. In cylindrical coordinates, the EOM (\ref{e:EOM-EM-component-form}) become
	\beq
	\mu \ddot{r} = \mu r \dot{\tht}^2 - q k^2 r + \frac{q}{c} ( \la k r \dot z - \la k m r \dot{\tht} ), \qquad
	\mu r \ddot{\tht} = \frac{q}{c} \la k m \dot{r} - 2 \mu \dot{r} \dot{\tht}  \quad \text{and} \quad 
	\mu \ddot{z} = -\frac{q}{c} \la k r \dot{r}.
	\eeq
	
%---------------------------------------------	
\subsection{Quantization of the electromagnetic Hamiltonian}
\label{s:Quantization-of-the-electromagnetic-Hamiltonian}	
%---------------------------------------------	

To quantize in Cartesian coordinates, we represent the canonical momenta as $p_x = -i \hbar \pdr_x$ etc., satisfying the canonical commutation relations $[ x, p_x ] = i \hbar$ etc. The Hamiltonian (\ref{e:Hamiltonian-EM-Cartesian}) becomes
	\beqs
	\hat{H} &=& \frac{1}{2 \mu} \left[ \left(  -i \hbar \pdr_x - \frac{q\la k m y}{2c} \right)^2 + \left(  -i \hbar \pdr_y - \frac{q\la k m x}{2c} \right)^2 + \left(  -i \hbar \pdr_z - \frac{q\la k (x^2 + y^2)}{2c} \right)^2 \right] \cr
	&&+ \frac{q k^2}{2}(x^2 + y^2 + m^2).
	\eeqs
To facilitate separation of variables, we work in cylindrical coordinates, where
	\beq
	\pdr_x = \cos \tht \pdr_r - \frac{\sin \tht}{r} \pdr_\tht \quad \text{and} \quad \pdr_y = \sin \tht \pdr_r - \frac{\cos \tht}{r} \pdr_\tht.
	\eeq
Thus, we have
	\beq
	\hat{H} = -\frac{\hbar^2}{2 \mu } \ov{r} \pdr_r [ r \pdr_r] + \ov{2 \mu r^2} \left[-i \hbar \pdr_{\tht} - \frac{q A_{\tht}(r)}{c} \right]^2 + \frac{1}{2 \mu} \left[-i \hbar\pdr_z - \frac{q A_z(r)}{c} \right]^2  + q V(r),
	\label{e:Quantum-Hamiltonian}
	\eeq
with the potentials given in (\ref{e:canonical-components-of-electro-magnetic-potentials}). We now introduce the momentum operators
	\beq
	\hat p_r = -i \hbar \frac{1}{\sqrt{r}} \pdr_r \sqrt{r} = -i \hbar \left( \pdr_r +  \ov{2 r}  \right), \quad \hat p_{\tht} = - i \hbar \pdr_{\tht} \quad \text{and} \quad  \hat p_z = -i \hbar \pdr_z.
	\label{e:canonical-momenta-cylindrical-coordinates}
	\eeq
They furnish a representation of the canonical commutation relations $[r, p_r] = i \hbar$ etc., and are Hermitian relative to the inner product $\bra \phi |  \psi  \ket = \int r dr d\tht dz \:  \phi^* \psi$. The Hamiltonian (\ref{e:Quantum-Hamiltonian}) becomes:
	\beq
	\hat{H} = \frac{1}{2 \mu} \left[ \hat p_r^2 +  \frac{1}{r^2} \left(\left( \hat p_{\tht} - \frac{q A_\tht}{c} \right)^2 - \frac{\hbar^2}{4} \right)+ \left(\hat p_z - \frac{q A_z}{c} \right)^2 \right] + q V(r).
	\label{e:quantum-Hamiltonian-cylindrical}
	\eeq
Here, $\hat p_r^2 = -\hbar^2 ( \pdr_{r}^2 + (1/r) \pdr_r - 1/4r^2)$. This Hamiltonian differs from the direct quantization of the classical cylindrical Hamiltonian (\ref{e:Hamiltonian-EM-plane-polar}) by a centripetal potential $-\hbar^2/ 8 \mu r^2$. Thus, we have chosen to define the quantum theory via canonical quantization in Cartesian coordinates.  

We can separate variables in the Schr\"odinger equation using the symmetries of (\ref{e:quantum-Hamiltonian-cylindrical}). The potentials $A_\tht$, $A_z$ and $V(r)$ are independent of $z$ and $\tht$ so that $\hat H, \hat{p}_z = -i \hbar \pdr_z$ and $\hat{p}_\tht = -i \hbar \pdr_\tht $ can be chosen to have common eigenstates. This leads to the factorized energy eigenfunctions:
\beq
	\psi(r, \tht, z) = \frac{1}{\sqrt{r}} \varrho(r) \exp(i l \tht) \exp \left(\frac{i p_z z}{\hbar} \right),
	\label{e:psi-seperation-of-variables}
	\eeq
where $p_z$ is a real number and $l$ must be an integer on account of the $2 \pi$-periodicity of $\tht$. Putting $\hat{H} \psi = E \psi$ we get the radial eigenvalue problem
	\beq
	-\frac{\hbar^2 \varrho''(r)}{2 \mu} + U_{\rm eff}(r) \varrho(r)  = E \varrho(r).
	\label{e:eigenvalue-problem}	
	\eeq
The $1/ \sqrt{r}$ prefactor in (\ref{e:psi-seperation-of-variables}) eliminates the $\varrho'$ that arises from the operator $\hat p_r^2$. Here, the effective potential
	\small
	\beqs
	U_{\rm eff}(r) &=& -\frac{\hbar^2}{8 \mu r^2} + \frac{1}{2\mu r^2} \left[ \hbar l - \frac{q A_{\tht}}{c} \right]^2 + \frac{1}{2 \mu} \left[ p_z - \frac{q A_z}{c} \right]^2 + q V(r) \cr
	       &=& \half \left[ \frac{\hbar^2\left[ l^2 - \frac{1}{4}\right]}{\mu r^2} + \frac{q \la k m \hbar l}{\mu c} + \frac{p_z^2}{\mu} + q k^2 m^2  + \left( \frac{ q^2 \la^2 k^2 m^2}{4 \mu c^2} - \frac{q \la k  p_z}{\mu c} + q k^2 \right) r^2  + \frac{q^2 \la^2 k^2 r^4}{4 \mu c^2} \right] \qquad
	\label{e:effective-potential-radial-part}
	\eeqs
	\normalsize	
includes inverse-square (attractive only if $l=0$), quadratic and quartic terms in $r$. 

This effective potential (in units where $\mu = q = c =1$) differs from that obtained by Rajeev and Ranken (in Eq. 4.8 of \cite{R-R}). In their expression for $U_{\rm eff}(r)$, $A_\tht$ was mistakenly taken as $\la m k r/2$ instead of $-\la m k r^2/2$. Thus, the corresponding radial equation they obtained in the strong coupling limit and the subsequent analysis to obtain the dispersion relation for quantized screwons needs to be reconsidered. In addition, they suggested that the strong coupling limit of the scalar field theory could also be interpreted as a `slow-light' post-relativistic regime. However, as we pointed out in \cite{G-V-1}, the ‘slow-light’ limit ($c \to 0$) holding $\la$ fixed is not quite the same as the strong-coupling limit of the scalar field theory. 

To be doubly sure of (\ref{e:effective-potential-radial-part}), we re-derive (\ref{e:eigenvalue-problem}) and (\ref{e:effective-potential-radial-part}) by interpreting the RR model as a quartic oscillator. This viewpoint will also facilitate the study of its quantum theory.

%------------------------------------------
\section{Rajeev-Ranken model as a quartic oscillator}
\label{s:Mechanical-interpretation-of-the-Rajeev-Ranken-Model}
%-------------------------------------------

Here, we interpret the Hamiltonian of the Rajeev-Ranken model (see Eqns.~(\ref{e:Hamiltonian-mech-Darboux}) and (\ref{e:Hamiltonian-EM-Cartesian}))
	\beqs
	 H &=& \half \left[ \left( k P_1 - \frac{\la k m R_2}{2} \right)^2 + \left( k P_2 + \frac{\la k m R_1}{2} \right)^2 + \left( k P_3 - \frac{\la k}{2}(R_1^2 + R_2^2) \right)^2 \right] \cr
	 && + \frac{k^2}{2}(R_1^2 + R_2^2 + m^2),
	\label{e:Hamiltonian-mech-RP}
	\eeqs
as that of a particle of mass $\mu = 1$, moving in a cylindrically symmetric quadratic plus quartic potential with an additional rotational energy. Indeed, taking the Darboux coordinates $R_{1, 2, 3}$ and their conjugate momenta $k P_{1,2,3}$ as the Cartesian components of the position and momentum of a particle of mass $\mu = 1$, (\ref{e:Hamiltonian-mech-RP}) becomes
	\beqs
	H &=& \frac{p_x^2 + p_y^2 + p_z^2}{2\mu} + \frac{\la k m (x p_y - y p_x)}{2\mu} + \left( \frac{\la^2 k^2 m^2 }{8 \mu} - \frac{\la k p_z}{2\mu} + \frac{k^2}{2} \right) (x^2 + y^2) \cr
	&& + \frac{\la^2 k^2}{8 \mu} (x^2 + y^2)^2 + \frac{k^2 m^2}{2}.
	\label{e:Hamiltonian-quadratic-quartic-Cartesian}
	\eeqs
The second term in $H$ is proportional to the angular momentum component $L_z$. To exploit the cylindrical symmetry of $H$ we switch to cylindrical coordinates defined in $\S \ref{s:Classical-charged-particle-in-an-axisymmetric-field}$ and obtain
	\beq
	H = \frac{1}{2 \mu} \left[ p_r^2 +  \frac{p_{\tht}^2}{r^2} + p_z^2 \right] + \frac{\la k m }{2 \mu} p_\tht  + \left( \frac{\la^2 k^2 m^2 }{8 \mu} - \frac{\la k p_z}{2 \mu} + \frac{k^2}{2} \right) r^2 + \frac{\la^2 k^2}{8 \mu} r^4 + \frac{k^2 m^2}{2}.
	\label{e:Hamiltonian-quadratic-quartic-cylindrical}
	\eeq
Although the terms linear in $p_{\tht}$ and $p_z$ are unconventional, the RR model requires them. Notice that the coupling constant $\la$ appears in both the quadratic and quartic coefficients as does the wavenumber $k$ of the screwons. When $k=0$, $H$ describes a free particle, while for $\la = 0$ it describes a cylindrically symmetric {\it harmonic} oscillator. 

%---------------------------------------------
\vspace{0.25cm}
%---------------------------------------------
\small 

{\bf \fl Remark:} Interestingly, the Hamiltonian of a quartic anharmonic oscillator $H =(1/2) \left( p^2 + \om^2 q^2 \right) + \la q^4$ can be re-expressed as a {\it quadratic} Hamiltonian by introducing the new variable $Q = q^2$: $H = (1/2) \left( p^2 + \om^2 q^2 \right) + \la Q^2$. However, unlike the step-2 nilpotent $q$-$p$ Heisenberg algebra, $Q$ and $p$ satisfy a step-3 nilpotent algebra: 
	\beq
	\{Q, p \} = 2q, \quad \{q, p \} = 1 \quad \text{and} \quad \{q, Q\} = 0.
	\eeq
Similarly, introducing $X = x^2$ and $Y = y^2$, we rewrite (\ref{e:Hamiltonian-quadratic-quartic-Cartesian}) as a quadratic Hamiltonian: 
	\beqs
	H &=& \frac{p_x^2 + p_y^2 + p_z^2}{2\mu} + \frac{\la k m (x p_y - y p_x)}{2\mu} + \left( \frac{\la^2 k^2 m^2}{8 \mu}+ \frac{k^2}{2} \right) (x^2 + y^2) \cr && - \frac{\la k p_z}{2\mu} (X + Y) + \frac{\la^2 k^2}{8 \mu} (X + Y)^2 + \frac{k^2 m^2}{2}.
	\eeqs
The EOM follow from this quadratic form on the step-3 nilpotent algebra:
	\beqs
	\{x, p_x \} &=&1,\quad  \{ X, p_x \} = 2 x, \quad \{ x, X \} = 0, \quad  \{ y, p_y \} = 1, \quad \{ Y, p_y \} = 2y, \cr 
	\{ y, Y \} &=& 0 \quad \text{and} \quad \{z, p_z \} = 1.
	\eeqs
%---------------------------------------------
\begin{comment}
Similarly, by relabelling $r^2 = s$, we may rewrite the cylindrical coordinate Hamiltonian (\ref{e:Hamiltonian-quadratic-quartic-cylindrical})
	\beq
	H = \frac{1}{2 \mu} \left[ p_r^2 +  \frac{p_{\tht}^2}{s} + p_z^2 \right] + \frac{\la m k}{2 \mu} p_\tht  + \left( \frac{\la^2 m^2 k^2}{8 \mu} - \frac{\la k p_z}{2 \mu} + \frac{k^2}{2} \right) s + \frac{\la^2 k^2}{8 \mu} s^2 + \frac{k^2 m^2}{2}.
	\eeq
The equations of motion follow from a step-3 nilpotent Poisson algebra among $p_r, r$ and $s$ and Heisenberg algebras among $\tht, p_{\tht}$ and $z, p_z$.
\end{comment}
%---------------------------------------------
This is similar to the formulation of the RR model in terms of  the variables (see \cite{R-R, G-V-1, G-V-2} and \S \ref{s:unitary-representation-of-nilpotent-Lie-algebra})
	\beq
	L = [K, R] + m K \quad \text{and} \quad S = \dot R + \frac{K}{\la},
	\label{e:L-and-S-variables}
	\eeq
where the Hamiltonian (\ref{e:Hamiltonian-RR-LS-variables-quantum}) is a quadratic form on a step-3 nilpotent Lie algebra. In this sense, the RR model joins the harmonic and anharmonic oscillators and Maxwell and Yang-Mills theories in their formulation in terms of  quadratic Hamiltonians on nilpotent Lie algebras. As mentioned in \cite{R-R}, this formulation may facilitate finding the spectrum of the Hamiltonian using the representation theory of the underlying nilpotent group \cite{J-K}. 

\normalsize

%---------------------------------------------
\vspace{0.25cm}
%---------------------------------------------

{\fl \bf Dimensional analysis:} Requiring that $H, p_{x,y,z}$ and $x, y, z$ have dimensions of energy, momentum and length, we find that the parameters in (\ref{e:Hamiltonian-quadratic-quartic-Cartesian}) have the following dimensions:
	\beq
	[\mu] = M, \quad [k] = M^{1/2} T^{-1}, \quad  [m] = L  \quad \text{and} \quad [\la] = M^{1/2} L^{-1}.
	\label{e:dimensions-of-parameters}
	\eeq
(This assignment of dimensions differs from that following from the relativistic scalar field theory \cite{R-R}, where $m, R, P$ are dimensionless while $[k]_{RR} = L^{-1} \quad \text{and} \quad [H]_{RR}= L^{-2}$). In particular, in the classical theory, $\tl \la= \la m/ \sqrt{\mu}$ is the only independent dimensionless combination and defines a nondimensional coupling constant. Since $p_z$ and $L_z$ are conserved quantities, from the structure of (\ref{e:Hamiltonian-quadratic-quartic-Cartesian}), the energy of any classical state can be expressed as 
	\beq
	E = \frac{p_z^2}{2 \mu} + \frac{\la k m L_z}{\mu} + m^2 k^2 f(\tl \la, \tl p_z, \tl L_z), \quad \text{where} \quad  \tl p_z = \frac{p_z}{k m \sqrt{\mu}} \quad \text{and} \quad  \tl L_z = \frac{L_z}{km^2 \sqrt{\mu}}.
	\label{e:classical-energy-dependence}
	\eeq
Here, $f$ is some function of the three dimensionless variables $\tl \la, \tl p_z $ and $\tl L_z$.

%---------------------------
\section{Quantum Rajeev-Ranken model}
\label{s:Quantum-RR-model}
%---------------------------

Even before formally quantizing the RR model, we may infer the possible dependence of energy eigenvalues on parameters from dimensional analysis. From (\ref{e:dimensions-of-parameters}), since $k m^2 \sqrt{\mu}$ has dimensions of action, in the quantum theory $\tl \hbar = \hbar/ k m^2 \sqrt{\mu}$ is a second independent dimensionless combination in addition to the classically present dimensionless coupling constant $\tl \la = \la m/ \sqrt{\mu}$. Thus, generalizing (\ref{e:classical-energy-dependence}), the energy of any quantum state must be of the form
	\beq
	E = \frac{p_z^2}{2 \mu} + \frac{\la k m L_z}{\mu} + m^2 k^2 g\left(\tl \la, \tl \hbar,  \frac{m p_z}{\hbar}, \frac{L_z}{\hbar} \right), 
	\label{e:spectrum-quantum-RR}
	\eeq
for some function $g$ of four conveniently chosen dimensionless combinations.

We will now quantize the RR model as an isotropic anharmonic oscillator. Quantum anharmonic oscillators have been studied in various contexts and several results are available. For instance, the Schr\"odinger eigenvalue problem for the one-dimensional (1D) quartic oscillator may be reduced  \cite{D-S-A-C-D} to the triconfluent Heun equation ($[0, 0, 1_6]$ in Ince's classification, see Appendix \ref{A:Singularities-of-second-order-ordinary-differential-equations}). The energy levels of this oscillator display remarkable analytic properties in the complex coupling constant plane \cite{B-W}.  Some exact results are available for the $N$-dimensional isotropic sextic oscillator \cite{D-W}, but they do not extend to the quartic version. Hill determinants have been used to numerically obtain the spectrum of 1D anharmonic oscillators \cite{B-D-S-S-V} as well as 2D isotropic quartic oscillators by truncating a Frobenius series expansion \cite{Taseli}. In the semi-classical regime, there are results on exact WKB quantization conditions and resurgent expansions for the spectrum of quartic oscillators \cite{Z-J-1,Z-J-2, Voros} (see \S \ref{s:Discussion} for more on this).

%---------------------------------------------
\vspace{0.25cm}
%---------------------------------------------

{\fl \bf Canonical quantization:} Upon quantization in Cartesian coordinates ($\hat p_x = -i \hbar \pdr_x$ etc.), the Hamiltonian (\ref{e:Hamiltonian-quadratic-quartic-Cartesian}) becomes
	\beq
	\hat H = \half \left[ \frac{\hat p_x^2 + \hat p_y^2 + \hat p_z^2}{\mu} + \frac{\la k m \hat L_z}{\mu} + \left( \frac{\la^2 k^2 m^2}{4 \mu} - \frac{\la k \hat p_z}{\mu} + k^2 \right) r^2 + \frac{\la^2 k^2}{4 \mu} r^4 + k^2 m^2 \right].
	\label{e:quantum-Hamiltonian}
	\eeq
Here, $ \hat L_z = x \hat p_y - y \hat p_x$ and $\hat p_z$ are conserved and $r^2 = x^2 + y^2$. To facilitate separation of variables, we switch to cylindrical coordinates (\ref{e:canonical-momenta-cylindrical-coordinates}), in which (\ref{e:quantum-Hamiltonian}) becomes:
	\beq
	\hat{H} = \frac{1}{2 \mu} \left[ \hat p_r^2 +  \frac{ \hat p_{\tht}^2 - \frac{\hbar^2}{4}}{r^2} + \hat p_z^2 \right] + \frac{\la k m}{2 \mu} \hat p_\tht  + \left( \frac{\la^2 k^2 m^2}{8 \mu} - \frac{\la k \hat p_z}{2 \mu} + \frac{k^2}{2} \right) r^2 + \frac{\la^2 k^2}{8 \mu} r^4 + \frac{k^2 m^2}{2}.
	\label{e:quantum-Hamiltonian-cylindrical-quadratic-quartic}
	\eeq
As before, we may write the energy eigenfunctions as
	\beq
	\psi = \rho(r) \exp(i l \tht) \exp \left( \frac{i p_z z}{\hbar} \right), \quad \text{where} \quad l \in \mathbb{Z}.
	\label{e:sep-of-var-common-eigenstate}
	\eeq
Separating variables in $\hat H \psi = E \psi$, we arrive at the radial eigenvalue problem
	\beq
	-\frac{\hbar^2}{2 \mu} \left( \rho''(r) + \ov{r} \rho'(r) - \frac{l^2}{r^2} \rho(r)  \right) +  U(r) \rho = \left( E -\frac{p_z^2}{2 \mu} - \frac{\hbar l \la k m}{2 \mu} - \frac{k^2 m^2}{2}  \right) \rho,
	\label{e:radial-eigenvalue-problem-for-quad-quartic}
	\eeq
with  the potential
	\beq
	U(r) = \al r^2 + \beta r^4 \quad \text{where} \quad \al = \frac{\la^2 k^2 m^2}{8 \mu} - \frac{\la k p_z}{2 \mu} + \frac{k^2}{2} \quad \text{and} \quad \beta = \frac{\la^2 k^2}{8 \mu}.
	\label{e:potential-radial-quad-quartic}
	\eeq
If instead of (\ref{e:sep-of-var-common-eigenstate}), we use the wave function (\ref{e:psi-seperation-of-variables}), then the resulting potential (in units where $q = c = 1$) agrees with the effective potential (\ref{e:effective-potential-radial-part}), which was obtained using the electromagnetic interpretation of \S \ref{s:Electromagnetic-interpretation-of-the-RR-model}. This confirms that the effective potential in \cite{R-R} is incorrect.

When $k = 0$, the potential $U(r)$ is absent and (\ref{e:radial-eigenvalue-problem-for-quad-quartic}) reduces to the Bessel equation \cite{B-O}. In this case, $E - p_z^2/2 \mu$ is the energy of a free particle on the $x$-$y$ plane, so it must be $ \geq 0$ irrespective of the value of $l$. More generally, it is convenient to separate out the free particle motion in the $z$-direction and define the 2D isotropic anharmonic oscillator Hamiltonian
	\beq
	\hat H_1 = \hat H -\frac{p_z^2}{2 \mu} - \frac{k^2 m^2}{2} = \frac{1}{2 \mu} \left( \hat p_r^2 + \frac{\hat p_\tht^2 - \frac{\hbar^2}{4}}{r^2} \right) + \frac{\la k m}{2 \mu} \hat p_{\tht} + U(r) ,
	\label{e:Hamiltonian-two-dimensional-anharmonic-oscillator}
	\eeq
with the eigenvalue 
	\beq
	E_1 = E - \frac{p_z^2}{2 \mu} - \frac{k^2 m^2}{2}.
	\label{e:shifted-energy-eigenvalue}
	\eeq
The coefficient $\beta$ of $r^4$ in (\ref{e:potential-radial-quad-quartic}) is positive while that of $r^2$ ($\alpha$) can have either sign. Thus, $U(x,y)$ is either convex or shaped like a Mexican-hat. In either case, (\ref{e:quantum-Hamiltonian}) reveals that the spectrum of $\hat H_1$ is bounded below and discrete (for any fixed value of $p_z$).

%----------------------------
\vspace{0.25cm}
%----------------------------

{\fl \bf Goldstone mode of the RR model:} When $\al < 0$, one could mistakenly treat minima of the Mexican-hat potential (\ref{e:potential-radial-quad-quartic}) as static solutions of the anharmonic oscillator. This is incorrect because of the term proportional to $L_z$ in (\ref{e:quantum-Hamiltonian}). The true static solutions are given by the solutions of the EOM (here $x,y,z = R_{1,2,3}$ and $p_{x,y,z} = kP_{1,2,3}$):
	\small
	\beqs
	\dot{x} &=& p_x - \frac{\la k m y}{2}, \quad \dot{y} = p_y + \frac{\la k m x}{2}, \quad \dot{z} = p_z - \frac{\la k}{2}(x^2 + y^2), \cr
	\dot{p_x} &=& -\frac{\la k m p_y}{2} - \left( \frac{\la^2 k^2 m^2}{8} -\frac{\la k p_z}{2} + \frac{k^2}{2} \right) 2 x -\frac{\la^2 k^2}{2}(x^2 + y^2) x \cr
	\dot{p_y} &=& \frac{\la k m p_x}{2} - \left( \frac{\la^2 k^2 m^2}{8} -\frac{\la k p_z}{2} + \frac{k^2}{2} \right) 2 y -\frac{\la^2 k^2}{2}(x^2 + y^2) y \quad \text{and} \quad \dot{p_z} = 0,
	\eeqs
	\normalsize
with $\dot R_{1,2,3} = 0$ and $\dot P_{1,2,3} =0$. It is possible to show that there is a one parameter family of static solutions parametrized by arbitrary real values of $R_3(t) = R_3$, while the other variables vanish: $R_{1,2} = P_{1,2,3} = 0$. [In terms of the $L$-$S$ variables of Eqn.~(\ref{e:L-and-S-variables}), this corresponds to a single point on the static submanifold $\Sigma_2$ (defined by $S_{1,2} = L_{1,2} =0$ in \S5.5 of \cite{G-V-1}), where $L_3 = -m k$ and $S_3 = -k/\la$. This is because the $L$-$S$ phase space does not include the $R_3$ degree of freedom.] These static solutions lie on the $z$ axis and are degenerate in energy ($E= m^2 k^2 /2$). Thus, we would expect a zero mode/`Goldstone mode' where $R_3$ varies slowly, while the other variables remain zero. However, we do not expect Goldstone bosons in the parent scalar field theory since it is two-dimensional.

%---------------------------------------------
\vspace{0.25 cm}
%---------------------------------------------

{\bf \fl Strong coupling limit at high energies:} In the strong coupling limit, $\la \to \infty$, the coefficients in the potential $U$ (\ref{e:potential-radial-quad-quartic}) simplify and $\alpha \approx m^2 \beta$. Thus, the radial equation (\ref{e:radial-eigenvalue-problem-for-quad-quartic}) becomes 
	\beq
	-\frac{\hbar^2}{2 \mu} \left(  \frac{{\rm d^2}}{{\rm d}r^2}  + \frac{1}{r} \frac{{\rm d}}{{\rm d}r} - \frac{l^2}{r^2} \right) \rho(r)+ \frac{\la^2 k^2}{8 \mu} (m^2 r^2 + r^4)\rho = \left( E -\frac{p_z^2}{2 \mu} - \frac{\hbar l \la k m}{2 \mu} - \frac{k^2 m^2}{2}  \right) \rho.
	\label{e:rad-eqn-str-coupl-naive}
	\eeq
For highly excited states as $\la \to \infty$, we ignore the $p_z^2/2\mu$ and $k^2 m^2/2$ terms on the RHS:
	\beq
	-\frac{\hbar^2}{2 \mu} \left(  \frac{{\rm d^2}}{{\rm d}r^2}  + \frac{1}{r} \frac{{\rm d}}{{\rm d}r} - \frac{l^2}{r^2} \right) \rho(r)+ \frac{\la^2 k^2}{8 \mu} (m^2 r^2 + r^4)\rho \approx \left( E^{\rm high} - \frac{\hbar l \la k m}{2 \mu}  \right) \rho.
	\label{e:radial-equation-highly-excited}
	\eeq
(The corresponding strong coupling radial equation (4.11) in \cite{R-R} has an error, $r^2$ is missing from $m^2 r^2$.) Notice that $\la$ and $k$ appear only through their product $\la k$. Thus, on dimensional grounds, the energies of highly excited states must be of the form $E^{\rm high} \approx (\hbar^2/\mu m^2) \epsilon(\tl \sig, l)$, where $\epsilon$ is a function of the dimensionless variables $\tl \sig = \la k m^3/ \hbar$ and $l$. We will argue in \S \ref{s:Numerical-WKB-spectrum} that, for fixed $l$, in the semi-classical approximation, $\epsilon(\tl \sig, l) \propto \tl \sig^{2/3}$. 

In \S \ref{s:Weak-and-strong-coupling-limits-of-the-Schrodinger-eigenvalue-problem}, we will consider a weak coupling and a double-scaling strong coupling limit for which it is convenient to work with dimensionless variables. 

%---------------------------------------------
\subsection{Quantum RR model in terms of dimensionless variables} 
\label{s:quantum-RR-model-in-terms-of-dimensionless-variables}
%---------------------------------------------

Assuming $k, m \neq 0$  ($k = 0$ corresponds to a free particle), we may re-write the Hamiltonian (\ref{e:quantum-Hamiltonian}) in terms of the dimensionless variables:	
	\beq
	(\tl x, \tl y, \tl z) = \frac{1}{m}(x,y,z), \quad \tl p_{x,y,z} = \frac{p_{x,y,z}}{k m \sqrt{\mu}} = -i \tl \hbar \pdr_{\tl x, \tl y, \tl z}, \quad
	\tl \la = \frac{\la m}{\sqrt{\mu}} \quad \text{and} \quad  \tl \hbar = \frac{\hbar}{ k m^2 \sqrt{\mu}}.
	\label{e:definitions-dimensionless-variables}
	\eeq
Dividing (\ref{e:quantum-Hamiltonian}) by $k^2 m^2/2$ we get the dimensionless Hamiltonian 
	\beq
	\tl H = \tl p_x^2 + \tl p_y^2 + \tl p_z^2 + \tl \la \tl L_z + \left( \frac{\tl \la^2}{4} - \tl \la \tl p_z + 1 \right)(\tl x^2 + \tl y^2) + \frac{\tl \la^2}{4}(\tl x^2 + \tl y^2)^2 + 1.
	\label{e:Hamiltonian-dimensionless-Cartesian}
	\eeq
Here $\tl L_z  = \tl x \tl p_y - \tl y \tl p_x$. Similarly, if $\tl r = r/m$, the cylindrical Hamiltonian (\ref{e:quantum-Hamiltonian-cylindrical-quadratic-quartic}) becomes
	\beq
	\tl H = -\tl \hbar^2 \left[  \frac{\pdr^2}{\pdr \tl r^2} + \ov{\tl r} \dd{}{\tl r}  + \frac{1}{\tl r^2} \frac{\pdr^2}{\pdr \tht^2} +  \frac{\pdr^2}{\pdr \tl z^2}\right] - i  \tl \hbar \tl \la \frac{\pdr}{\pdr \tht} + \left(\frac{\tl \la^2}{4} + i \tl \hbar \tl \la \frac{\pdr}{\pdr \tl z} + 1 \right) \tl r^2 + \frac{\tl \la^2 }{4} \tl r^4 + 1.
	\label{e:Hamiltonian-dimensionless-cylindrical-coordinates}
	\eeq
The use of dimensionless couplings will facilitate taking strong and weak coupling limits in \S \ref{s:Weak-and-strong-coupling-limits-of-the-Schrodinger-eigenvalue-problem}. 

As before, we separate variables in $\tl H \psi = \tl E \psi$  (where $\tl E = 2 E/k^2 m^2$) by putting $\psi = \rho(\tl r) \exp(i l \tht) \exp\left( i \tl p_z \tl z/\tl \hbar \right)$, to arrive at the radial equation
	\beq
	-\tl \hbar^2 \left( \rho''(\tl r) + \ov{\tl r} \rho'(\tl r) - \frac{l^2}{\tl r^2} \rho(\tl r)  \right) +  \tl U( \tl r) \rho = \left( \tl E - \tl p_z^2 - l \tl \hbar \tl \la - 1  \right) \rho,
	\label{e:eigen-value-problem-cylindrical-coordinates-dimensionless}
	\eeq
with  the potential 
	\beq
	\tl U(\tl r) = \tl \al \tl r^2 + \tl \beta \tl r^4 \quad \text{where} \quad \tl \al = \frac{\tl \la^2}{4} - \tl \la \tl p_z + 1 = \frac{2 \al}{k^2} \quad \text{and} \quad \tl \beta = \frac{\tl \la^2}{4} = \frac{2 \beta m^2}{k^2}.
	\eeq
As in (\ref{e:Hamiltonian-two-dimensional-anharmonic-oscillator}), we define a dimensionless Hamiltonian for a 2D anharmoic oscillator:
	\beq
	\tl{H_1} = \tl H - \tl p_z^2 - 1 = -\tl \hbar^2 \left[ \frac{\pdr^2}{\pdr \tl r^2} + \ov{\tl r} \dd{}{\tl r}  + \frac{1}{\tl r^2} \frac{\pdr^2}{\pdr \tht^2} \right] -i \tl \hbar \tl \la \frac{\pdr}{\pdr \tht} + \tl U(\tl r).
	\label{e:Hamiltonian-dimensionless-2-dimension}
	\eeq
Thus, the radial equation can be rewritten as 
	\beq
	-\tl \hbar^2 \left( \rho''(\tl r) + \ov{\tl r} \rho'(\tl r) - \frac{l^2}{\tl r^2} \rho(\tl r)  \right) + (\tl U(\tl r) + l \tl \hbar \tl \la) \rho = \tl E_1 \rho \quad \text{where} \quad \tl E_1 = \tl E - \tl p_z^2 - 1.
	\label{e:radial-equation-dimensionless-coupling}
	\eeq

%----------------------	
\vspace{0.25cm}	
%----------------------
	
{\bf \fl Normalizability condition:} From the inner product $\bra \phi |  \psi  \ket = \int \phi^* \psi \; r \: dr \: d\tht \: dz$, we get the normalizability condition for radial bound states: 
	\beq
	\bra \rho |  \rho  \ket = m^2\int \tl r \rho^2(\tl r) \:  d\tl r < \infty.
	\label{e:normalizability-condition-radial-bound-state}
	\eeq
Thus, $\rho(\tl r)$ must decay faster than $1/\tl r$ as $\tl r \to \infty$ and grow slower than $1/ \tl r$ as $\tl r \to 0$.

%---------------------------------------------
\subsection{Weak and double-scaling strong coupling limits}
\label{s:Weak-and-strong-coupling-limits-of-the-Schrodinger-eigenvalue-problem}
%---------------------------------------------

As will be discussed in \S \ref{s:Properties-of-the-radial-Schrodinger-equation}, the radial equations (\ref{e:radial-eigenvalue-problem-for-quad-quartic}) and (\ref{e:radial-equation-dimensionless-coupling}) are not solvable in terms of Lam\'e, Heun or other familiar functions. Here, we consider a weak coupling and a double-scaling strong coupling limit. The energy spectrum in the weak coupling limit is obtained. In the double-scaling strong coupling limit, we find the dependence of energy levels on the wavenumber $k$. These results are used to deduce dispersion relations for quantized screwons in these limits. 

%---------------------------------------------
\vspace{0.25cm}
%---------------------------------------------

{\fl \bf Weak coupling limit:} When $\tl \la \to 0$, (\ref{e:Hamiltonian-dimensionless-Cartesian}) reduces to $\tl H_{\tl \la \to 0} = \tl p_x^2 + \tl p_y^2 + \tl p_z^2 + \tl x^2 + \tl y^2 + 1$. Omitting the free particle motion in the $z$-direction, we arrive at a 2D harmonic oscillator $\tl H_{1_{\tl \la \to 0}} = \tl p_x^2 + \tl p_y^2 + \tl x^2 + \tl y^2$. The radial equation (\ref{e:eigen-value-problem-cylindrical-coordinates-dimensionless}) reduces to a confluent hypergeometric equation
	\beq
	-\tl \hbar^2 \left( \rho''(\tl r) + \ov{\tl r} \rho'(\tl r) - \frac{l^2}{\tl r^2} \rho(\tl r)  \right) + \tl r^2 \rho = \tl E_{1_{\tl \la \to 0}} \rho.
	\label{e:weak-coupling-radial-equation}
	\eeq
The spectrum is given by
	\beq
	\tl E_{1_{\tl \la \to 0}} = \lim_{\tl \la \to 0} (\tl E - \tl p_z^2 - 1) =  2 |\tl \hbar| (n_x + n_y + 1) \quad \text{with} \quad n_x, n_y = 0,1,2, \ldots.
	\eeq
In terms of the principal quantum number $n = 0,1,2, \ldots$, $n_x + n_y = 2 n + |l|$, where $l \in \mathbb{Z}$, with degeneracies given by $n_x+n_y+1$ \cite{Pauli}. Re-instating dimensions, we get the weak coupling spectrum of the RR model:
	\beq
	\lim_{\la \to 0} E =  \frac{k^2 m^2}{2} + (n_x + n_y + 1) \frac{\hbar \, |k|}{\sqrt{\mu}} + \frac{p_z^2}{2 \mu}, \quad \text{where} \quad n_x, n_y = 0,1,2, \ldots.
	\label{e:weak-coupling-dispersion-relation}
	\eeq
Since $k$ is the wavenumber of screwons, we see that quantized screwons at weak coupling display a linear dispersion relation typical of free relativistic particles. The quadratic term $k^2 m^2/2$ in (\ref{e:weak-coupling-dispersion-relation}) is a constant addition to the relativistic energy per unit length that arises when the ansatz $\phi = e^{Kx} R(t) e^{-Kx} + m K x$ is inserted in the field energy density $(1/2) (\dot \phi^2 + \phi'^2)$.

%Alternatively, when $\tl \la \to 0$, the radial equation (\ref{e:eigen-value-problem-cylindrical-coordinates-dimensionless}) 
%	\beq
%	-\tl \hbar^2 \left( \rho''(\tl r) + \ov{\tl r} \rho'(\tl r) - \frac{l^2}{\tl r^2} \rho(\tl r)  \right) + \tl r^2 \rho = \tl E_{1_{\tl \la \to 0}} \rho
%	\label{e:weak-coupling-radial-equation}
%	\eeq
%reduces to a confluent hypergeometric equation leading to the spectrum:
%	\beq
%	\tl E_{1_{\tl \la \to 0}} = 2 |\tl \hbar| ( 2n + |l| + 1), \quad \text{where} \quad 2 n + |l| = n_x + n_y.
%	\label{e:weak-coupling-dispersion-relation-hypergeometric}
%	\eeq
%---------------------------------------------

%---------------------------------------------
\vspace{0.25cm}
%---------------------------------------------

{\fl \bf Double-scaling strong coupling limit:} We now let $\tl \la, \tl \hbar \to \infty$, holding $\tl g = \tl \la/ \tl \hbar$ and $\tl p_z$ finite, so that all terms on the LHS of (\ref{e:radial-equation-dimensionless-coupling}) grow like $\tl \la^2$. To get a nontrivial eigenvalue problem, we focus on eigenvalues $\tl E_1$ that grow quadratically with $\tl \la$ and consequently define $\tl E_2 = \tl E_1  /\tl \la^2 $. Holding $\tl p_z$ finite ensures that the $k$-dependence drops out and the radial equation (\ref{e:radial-equation-dimensionless-coupling}) becomes:
	\beq
	 \rho''(\tl r) + \ov{\tl r} \rho'(\tl r) - \left( \frac{l^2}{\tl r^2} + \frac{\tl g^2}{4}\left(\tl r^2 + \tl r^4 \right)  + \tl g l  \right) \rho(\tl r) = - \tl g^2 \tl E_2 \, \rho(\tl r).
	 \label{e:strong-coupling-radial-equation-dimensionless}
	\eeq
The $\tl p_z$ terms drop out of the Hamiltonian (\ref{e:Hamiltonian-dimensionless-Cartesian}) in this double-scaling limit:
	\beq
	\frac{\tl H}{\tl \hbar^2}  = -\left(\frac{\pdr^2}{\pdr \tl x^2} + \frac{\pdr^2}{\pdr \tl y^2} \right)  - i \tl g \left( \tl x \frac{\pdr}{\pdr \tl y} - \tl y \frac{\pdr}{\pdr \tl x} \right) + \frac{\tl g^2}{4} ( \tl x^2 + \tl y^2 + ( \tl x^2 + \tl y^2 )^2).
	\eeq 
It follows that the finite rescaled energy eigenvalue in this limit $\tl E_2 (\tl g , l)$ is independent of $k$. Thus, in this strong coupling limit, the nontrivial dimensionless energy eigenvalues $\tl E \sim \tl \la^2 \tl E_2(\tl g, l) + \tl p_z^2 + 1$ must diverge quadratically in $\tl \la$ with the last two terms being sub-leading. Finally, the original (dimensionful) energy (see Eq.~(\ref{e:eigen-value-problem-cylindrical-coordinates-dimensionless})) $E$ is
	\beq
	E_{\text{ds-strong}} = \frac{k^2 m^2}{2} \tl E \sim \frac{k^2 m^2}{2} \tl \hbar^2 \left(  \tl g^2 \tl E_2 (\tl g, l)  + \frac{\tl p_z^2 + 1}{\tl \hbar^2}  \right) 
	= \frac{\hbar^2}{2 \mu m^2} \left( \tl g^2 \tl E_2 (\tl g, l) + \frac{\tl p_z^2 + 1}{\tl \hbar^2} \right).
	\eeq
Thus, in this double-scaling strong coupling limit ($\hbar, \la \to \infty$), the energy $E$ is quadratically divergent. The rescaled energy $E_{\text{ds-strong}}/\hbar^2 \propto k^0$, leading to a $k$-independent dispersion relation for quantized screwons in this limit.

Having arrived at the double-scaling strong coupling limit, we may send $\tl g \to 0$ resulting in a free particle moving on a plane. One must bear in mind that this double-scaling strong coupling limit with $\tl \hbar \to \infty$ is distinct from the naive strong coupling limit considered in (\ref{e:rad-eqn-str-coupl-naive}) and the strong coupling limit $(\tl \la \to \infty)$ of the classical model, where $\tl \hbar = 0$.

%---------------------------------------------
\vspace{0.25cm}
%---------------------------------------------

{\bf \fl Double-scaling limit in terms of $L$-$S$ variables:} In terms of $L$ and $S$(\ref{e:L-and-S-variables}), 
	\beq
	\tl p_z = \frac{p_z}{k m \sqrt{\mu}} = \frac{k P_3}{k m \sqrt{\mu}} =  \frac{\tl \la}{2 m^2}(2 \mathfrak{c} - m^2) + \frac{1}{ \tl \la \mu}.
	\eeq
Here, $\mathfrak{c} = (1/2)(L_1^2 + L_2^2 + L_3^2) + k S_3/\la$ and $m = -L_3/k$ are Casimirs of the $L$-$S$ Poisson algebra. To keep $\tl p_z$ finite in this strong coupling limit ($\tl \la \to \infty$), $\mathfrak{c}$ and $m$ must be chosen so that $2 \mathfrak{c} - m^2 \to 0$, in such a way that $\tl \la (2 \mathfrak{c} - m^2)$ approaches a finite limit. This effectively restricts the dynamics to a submanifold of the $L$-$S$ phase space on which the coordinates satisfy the relation $(L_1^2 + L_2^2)/k^2 = -2S_3/k \la$. Given $\mathfrak{c}$ and $m$, the dynamics is confined to a 3D submanifold labelled by $S_{1,2}$ and $r = \sqrt{L_1^2 + L_2^2}/k$.

%---------------------------------------------
\vspace{0.25cm}
%---------------------------------------------

{\bf \fl Double-scaling limit is anisotropic:} The strong coupling limit ($\tl \la , \tl \hbar \to \infty$, holding $\tl p_z$ fixed), is an anisotropic limit since the terms involving $\tl z$ in (\ref{e:Hamiltonian-dimensionless-cylindrical-coordinates}) are sub-leading compared to those involving $\tl x$ and $\tl y$. The Hamiltonian becomes
	\beq
	\tl H_2 = \tl H/ \tl \hbar^2 = \tl P_x^2 + \tl P_y^2 + \tl g \tl P_{\tht} + \frac{\tl g^2}{4} \left( \tl x^2 + \tl y^2 + (\tl x^2 + \tl y^2)^2 \right).
	\eeq
Here, $\tl P_x =  \tl p_x/ \tl \hbar= -i \pdr/ \pdr \tl x$, $\tl P_y = \tl p_y/ \tl \hbar = -i \pdr/ \pdr \tl y$ and $\tl P_{\tht} = \tl x \tl P_y - \tl y \tl P_x$ with the commutation relations $[ \tl x, \tl P_x ] = i $ and $[ \tl y, \tl P_y ] = i$. Thus, the double-scaling limit results in a dimensional reduction to a 2D quartic oscillator with an extra rotational energy. 

%---------------------------------------------
\subsection{Properties of the radial Schr\"odinger equation}
\label{s:Properties-of-the-radial-Schrodinger-equation}
%---------------------------------------------

The radial eigenvalue problems (\ref{e:radial-equation-dimensionless-coupling}), (\ref{e:radial-equation-highly-excited}) and  (\ref{e:strong-coupling-radial-equation-dimensionless}) are all of type $[0, 1, 1_6]$ in Ince's classification (see Appendix \ref{A:Singularities-of-second-order-ordinary-differential-equations}). This means they have a nonelementary regular singular point at $\tl r = 0$ and a rank 3 irregular singular point at $\tl r = \infty$.
%---------------------------------------------
\iffalse
Alternatively, to see this, we transform the independent variable  $\tl r = 1/s$, in terms of which (\ref{e:strong-coupling-radial-equation-dimensionless}) becomes (for simplicity we took the strong coupling radial equation):
	\beq
	\frac{d^2 \rho}{ds^2} + \frac{1}{s} \frac{d \rho}{ds} - \frac{1}{s^4} \left( l^2 s^2 + \frac{\tl g^2}{4}\left(\frac{1}{s^2} +  \frac{1}{s^4} \right)  - \tl g^2 \tl E_2 \right) \rho = 0.
	\eeq
Here, we have used the transformation rules
	\beq
	\frac{d \rho}{d \tl r} = -s^2 \frac{d \rho}{d s}  \quad \text{and}  \quad \frac{d^2 \rho}{d \tl r^2} = s^4 \frac{d^2 \rho}{ds^2}  + 2 s^3 \frac{d \rho}{ds}.
 	\eeq 
From this it is clear that $s = 0$ is a rank 3 irregular singular point ($K_1 = -1$ and $K_2 = 4$ from the order of poles at $\zeta = $0) and in turn $\tl r = \infty$ is an irregular singular point of (\ref{e:strong-coupling-radial-equation-dimensionless}). Thus the radial equation (\ref{e:strong-coupling-radial-equation-dimensionless}) has the designation $[0,1,1_6]$ which means the differential equation has no elementary singularities, one nonelementary regular singularity (at $\tl r = 0$) and an irregular singularity of species 6 (rank 3) at $\tl r = \infty$. 
\fi
%---------------------------------------------
The latter can be thought of as arising from a merger of four nonelementary regular singular points. Thus, we may regard our radial equations as confluent forms of an ODE with either 5 nonelementary or 10 elementary regular singular points (see \cite{Crowson} for an analysis of an ODE with 5 nonelementary singular points). In particular, our radial equations cannot be solved in terms of hypergeometric, Heun or Lam\'e functions or their confluent forms. 

By contrast, the weak coupling equation (\ref{e:weak-coupling-radial-equation}) is of type $[0,1, 1_4]$. The substitution  $\tl r^2 = x$ reduces the rank of its irregular singularity at $\tl r = \infty$  from 2 to 1, resulting in an equation of type $[0,1,1_2]$. The confluent hypergeometric equation is of the same type and (\ref{e:weak-coupling-radial-equation}) can be solved in terms of generalized Laguerre polynomials, which are special cases of the confluent hypergeometric function. 

Returning to the radial equation (\ref{e:radial-equation-dimensionless-coupling}), for large values of $\tl r$, the method of dominant balance \cite{B-O} (see Appendix \ref{B:Asymptotic-behaviour-of-the-radial-wavefunction}) leads to the asymptotic behaviour
	\beq
	\rho(\tl r) \sim \exp\left( -\frac{\sqrt{\tl \beta}}{\tl \hbar} \left( \frac{\tl r^3}{3} + \frac{ \tl \alpha \tl r}{2 \tl \beta} \right) \right) \tl r^{-3/2} a(\tl r), \quad \text{where} \quad a(\tl r) \sim \mathcal{O}(1) \quad \text{as} \quad \tl r \to \infty.
	\eeq 
The rank (three) of the singularity at $\infty$ determines the dominant asymptotic behaviour. In the double-scaling strong coupling limit of \S \ref{s:Weak-and-strong-coupling-limits-of-the-Schrodinger-eigenvalue-problem}, $\tl \al/ \tl \beta \to 1$ and $\sqrt{\tl \beta}/\tl \hbar  \to \tl g/2$. 

Around the regular singularity $\tl r = 0$ of (\ref{e:radial-equation-dimensionless-coupling}), the Frobenius series $\rho (\tl r) =  \tl r^\eta  \sum_{n = 0}^\infty \rho_n \tl r^n$ implies the exponents $\eta_{1,2} = \pm l$ (see (\ref{e:Indicial-equation-general-form})). The normalization condition (\ref{e:normalizability-condition-radial-bound-state}) picks out  $\eta = \eta_1$. In general, $\rho_n$ satisfy a four-term recurrence relation. The formulae are shorter in the double-scaling limit, where we get (see Appendix \ref{C:Frobenius-method-for-strong-coupling-limit-Local-analysis})
	\beqs
	\rho_1 &=& 0, \quad \rho_2 = \frac{- \tl g^2 \tl E_2 + l \tl g}{4 l + 4} \rho_0, \quad \rho_3 = 0, \quad (8 l + 16) \rho_4 = \frac{\tl g^2}{4} \rho_0 - (\tl g^2 \tl E_2 - l \tl g) \rho_2, \quad \rho_5 =0, \cr
	\text{and} &&(2 n l + n^2)\rho_n + (\tl g^2 \tl E_2 - l \tl g) \rho_{n-2} - \frac{\tl g^2}{4}(\rho_{n-4} - \rho_{n-6})  = 0,  \quad \text{for} \quad n = 6,8, \ldots,
	\label{e:recurrence-relations-around-zero-rho}
	\eeqs
with $\rho_{\rm odd} = 0$. By contrast, one has 2, 3 and 3-term recurrence relations for the hypergeometric, Lam\'e and Heun equations \cite{Ince, W-W}.

%-----------------------------------
\section{Semi-classical WKB approximation} 
\label{s:Semi-classical-WKB-approximation}
%-----------------------------------

In this section, we express the leading WKB quantization condition in terms of elliptic integrals. At weak coupling, the WKB spectrum agrees with (\ref{e:weak-coupling-dispersion-relation}). For larger $\la$, we have not been able to invert the elliptic integrals analytically but do so numerically to find the spectrum of quantized screwons.  

%---------------------------------------------
\subsection{WKB quantization condition}
\label{s:WKB-approximation}
%---------------------------------------------

Separating variables in cylindrical coordinates 
	\beq
	\psi(r, \tht, z) = \frac{\varrho(r)}{\sqrt{r}} \exp\left(\frac{i p_{\tht} \tht}{\hbar} \right) \exp \left(\frac{i p_z z}{\hbar}\right)
	\label{e:common-eigenstate}
	\eeq 
leads to the radial equation
\beq
	\frac{-\hbar^2}{2 \mu} \varrho'' + V_{\rm eff} \varrho = \left(E - \frac{p_z^2}{2 \mu} - \frac{p_{\tht} \la k m }{2 \mu} - \frac{k^2 m^2}{2} \right) \varrho, \quad \text{where} \quad V_{\rm eff} = U(r) + \frac{\sig}{r^2}
	\label{e:radial-equation-WKB}
	\eeq 
with $\sig = (\hbar^2/2\mu)( p_{\tht}^2/\hbar^2 - 1/4)$. Here, $U(r) = \al r^2 + \beta r^4$ is the potential from (\ref{e:potential-radial-quad-quartic}). Putting $\varrho(r) = \exp \left( i W(r)/\hbar \right)$ in (\ref{e:radial-equation-WKB}) gives
	\beq
	i \hbar W''(r) - W'(r)^2 + p(r)^2 = 0, \quad \text{where} \quad
	p(r)^2 = 2 \mu \left(E - \frac{p_z^2}{2 \mu} - \frac{p_{\tht} \la k m}{2 \mu} - \frac{k^2 m^2}{2}  - V_{\rm eff}\right).
	\label{e:radial-equation-WKB-and-classical-momentum}
	\eeq
We now expand $W$ and $E$ in the semi-classical series:
	\beq
	W(r) = W_0 + \hbar W_1+ \cdots \quad \text{and} \quad E = E^{(0)} + \hbar E^{(1)}+ \cdots.
	\eeq 
At $\mathcal{O}(\hbar^0)$, we obtain 
	\beq
	W_0(r) = \pm \int^r  dr \: \left[2 \mu E^{(0)} - p_z^2 - p_{\tht} \la k m - k^2 m^2 \mu - 2 \mu( \al r^2 + \beta r^4)  - \frac{p_{\tht}^2}{r^2} \right]^{1/2}, 
	\eeq
while at $\mathcal{O}(\hbar)$ we get $W_1' = ((iW_0''/2) + \mu E^{(1)})/W_0'$. Requiring $\varrho(r) \sim \exp \left( iW_0/ \hbar \right)$ to be singlevalued, we obtain the quantization condition
	\beq
	\int_{r_{\rm min}}^{r_{\rm max}} dr \, \left[2 \mu E^{(0)} - p_z^2 - p_{\tht} \la k m - k^2 m^2 \mu - 2 \mu( \al r^2 + \beta r^4)  - \frac{p_{\tht}^2}{r^2} \right]^{1/2} = n \pi \hbar,
	\label{e:quantization-condition}
	\eeq
where the radial quantum number $n$ is a (large) integer. Here, $r_{\rm min} < r_{\rm max}$ are the positive zeros of  $\lim_{\hbar \to 0} p(r)^2$ (\ref{e:radial-equation-WKB-and-classical-momentum}), enclosing the unique classically allowed interval. Similarly, for $\psi$ to be singlevalued on the $\tht$-circle, we must have $p_{\tht}= l \hbar$ for an integer $l$ of large magnitude. 

Using the substitution $x = r^2$, the quantization condition (\ref{e:quantization-condition}) becomes:
	\beq
	\int_{x_{\rm min}}^{x_{\rm max}} \frac{dx}{2x} \, \left[(2 \mu E^{(0)} - p_z^2 - p_{\tht} \la k m - k^2 m^2 \mu) x - 2 \mu( \al x^2 + \beta x^3)  - p_{\tht}^2 \right]^{1/2} = n \pi \hbar.
	\label{e:quantization-condition-cubic}
	\eeq
This integral can be expressed as a sum of complete elliptic integrals (see Appendix \ref{D:Computation-of-elliptic-integral}). However, we have not been able to invert it analytically except at weak coupling.

%---------------------------------------------
\vspace{0.25 cm}
%---------------------------------------------

{\fl \bf Weak coupling limit:}  When $\la \to 0$, $\al = k^2/2, \beta = 0$ and (\ref{e:quantization-condition}) simplifies. Putting $s = r^2$:
	\beq
	\int_{s_{\rm min}}^{s_{\rm max}} \frac{ds}{2 s} \sqrt{a s^2 + b s + c} = n \pi \hbar,
	\label{e:Integral-for-quatization-condition-isotropic-harmonic-oscillator-redefined}
	\eeq
where 
	\beq
	a = - k^2 \mu, \quad b = 2 \mu E_{1_{\la \to 0}} = 2 \mu \left(E^{(0)} - \frac{p_z^2}{2\mu} - \frac{k^2 m^2}{2}\right) \quad \text{and} \quad c = -p_{\tht}^2.
	\eeq 
Notice that  $a < 0$, $b > 0$ ($2 \mu \: \times$ energy of the 2D anharmonic oscillator at weak coupling) and $c < 0$. Using this, the turning points are:	
	\beq
	s_{\rm min, max} = \ov{k^2} \left[E_{1_{\la \to 0}} \mp \frac {\sqrt{\Delta}}{2 \mu}\right] > 0, \quad \text{where} \quad  \Delta = b^2 - 4 a c = 4 \mu^2 \left( E_{1_{\la \to 0}}^2 - \frac{k^2 p_{\tht}^2 }{\mu} \right). 
	\label{e:classical-turning-points}
	\eeq
The RHS of (\ref{e:classical-turning-points}) is to be interpreted in the classical limit, where commutators and other terms of $\mathcal{O}(\hbar)$ are ignored. We evaluate the LHS of (\ref{e:Integral-for-quatization-condition-isotropic-harmonic-oscillator-redefined}) using  Eq. 2.267(1) of \cite{G-R}:
%---------------------------------------------
\iffalse
	\beq
	\int \frac{\sqrt{R} \, dx}{x} = \sqrt{R} + a \int \frac{dx}{x \sqrt{R}} + \frac{b}{2} \int \frac{dx}{\sqrt{R}} \quad \text{where} \quad R = a + b x + c x^2.
	\label{e:indefinte-integral-square-root-quadratic-and-rational}
	\eeq 
For nonzero values of $l$, i.e. $c < 0, b < 0$ and $\Delta < 0$, we have 
	\beq	
	\int \frac{dx}{\sqrt{R}} = \frac{-1}{\sqrt{-c}} \arcsin \left( \frac{2 c x + b}{\sqrt{-\Delta}} \right) \quad \text{and} \quad
	\int \frac{dx}{x \sqrt{R}} = \ov{\sqrt{-a}} \arcsin \left( \frac{ 2 a + b x}{x \sqrt{-\Delta}} \right).
	\label{e:indefinte-integrals-inverse-trigonometric-functions}
	\eeq
\fi
%---------------------------------------------
	\beq
	\int_{s_{\rm min}}^{s_{\rm max}} \frac{ds}{2 s} \sqrt{a s^2 + b s + c} = \frac{\pi}{2} \left( \frac{\sqrt{\mu}}{|k|} E_{1_{\la \to 0}} - |p_{\tht}| \right).
	\eeq
This leads to the spectrum
	\beq
	E_{1_{\la \to 0}} \approx \left( 2n + \frac{|p_{\tht}|}{\hbar} \right) \frac{\hbar |k|}{\sqrt{\mu}} \quad \text{where} \quad p_{\tht} = l \hbar \quad \text{for} \quad l, n \gg 1.
	\label{e:weak-coupling-WKB-spectrum}
	\eeq
This weak coupling semiclassical result agrees with the exact weak coupling spectrum (\ref{e:weak-coupling-dispersion-relation}), if we identify $n_x + n_y$ with $2 n + |l|$ for $p_{\tht} / \hbar = l$ a large integer.

%------------------------------------
\subsection{Numerical WKB spectrum}
\label{s:Numerical-WKB-spectrum}
%--------------------

The WKB quantization condition (\ref{e:quantization-condition-cubic}) is 
	\beq
	\int_{x_{\rm min}}^{x_{\rm max}} \frac{dx}{2x} \, \left[(2 \mu E^{(0)} - p_z^2 - p_{\tht} \la k m - k^2 m^2 \mu) x - 2 \mu( \al x^2 + \beta x^3)  - p_{\tht}^2 \right]^{1/2} = n \pi \hbar
	\label{e:WKB-quantization-condition-cubic-polynomial}
	\eeq
where
	\beq
	\al = \frac{\la^2 k^2 m^2}{8 \mu} - \frac{\la k p_z}{2 \mu} + \frac{k^2}{2} \quad \text{and} \quad  \beta = \frac{\la^2 k^2}{8 \mu}.
	\label{e:alpha-and-beta-of-quartic-potential}
	\eeq
Although this integral is elliptic, we have not yet been able to invert it analytically to obtain the energy spectrum. Here, we do this numerically.

%--------------
Henceforth, we work in units where $\mu = m = \hbar = 1$. For fixed $k$ and $\la$, we evaluate the integral in (\ref{e:WKB-quantization-condition-cubic-polynomial}) numerically for a range of equally spaced $E^{(0)}$ values and interpolate between them. Requiring the resulting function of $E^{(0)}$ to equal an integer multiple of $\pi$, we extract the energy spectrum $E^{(0)}_n$ for several values of $k, \la, p_z,$ and $l$. The numerical scheme is validated using the known weak coupling spectrum and  displays the following features.

\begin{figure}[h]
		\begin{subfigure}[t]{5cm}
		\centering
		\includegraphics[width=6cm]{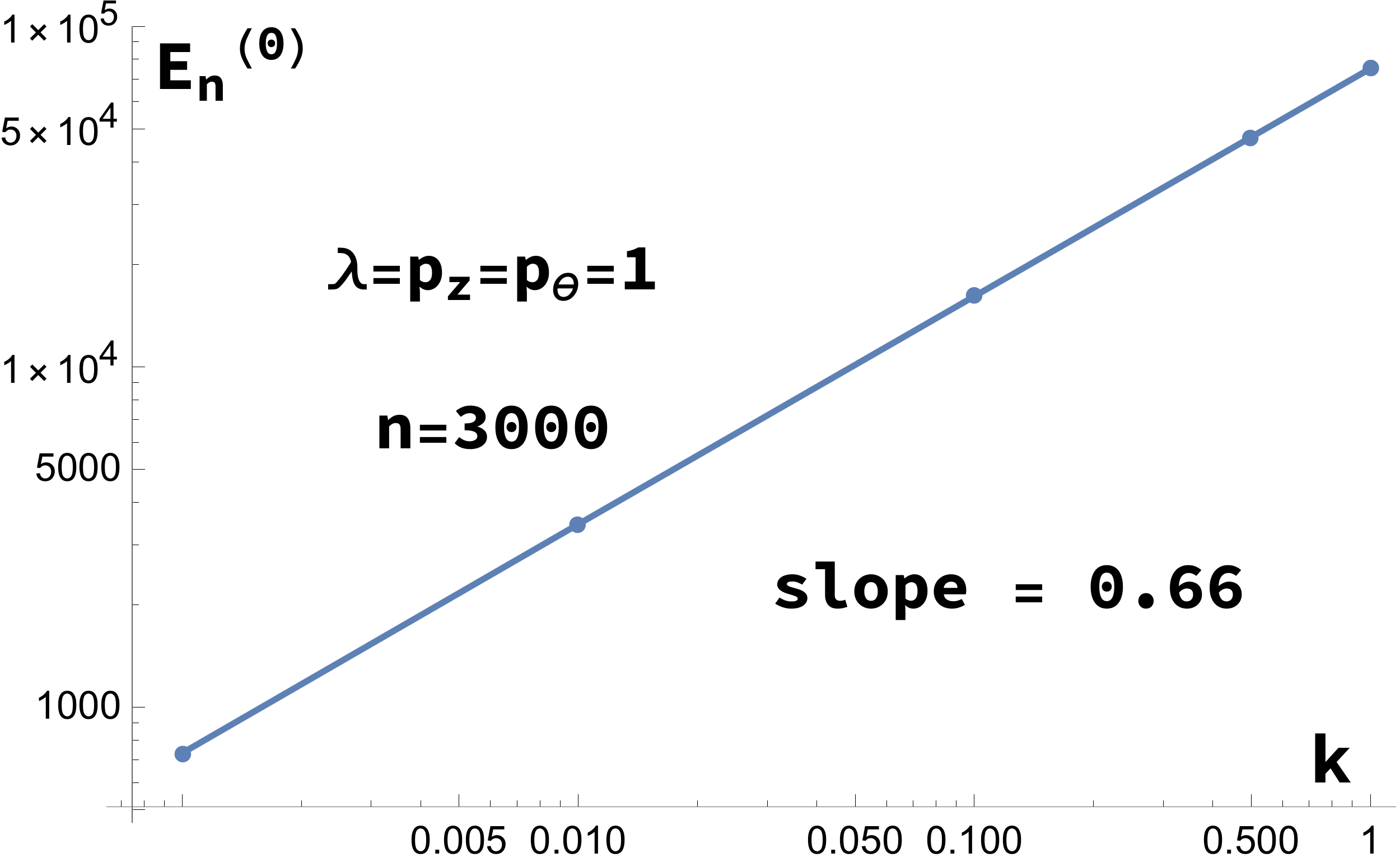}
		\caption{}
		\label{f:EvsK-la-1}
		\end{subfigure}
		\qquad
		\begin{subfigure}[t]{5cm}
		\centering
		\includegraphics[width=6cm]{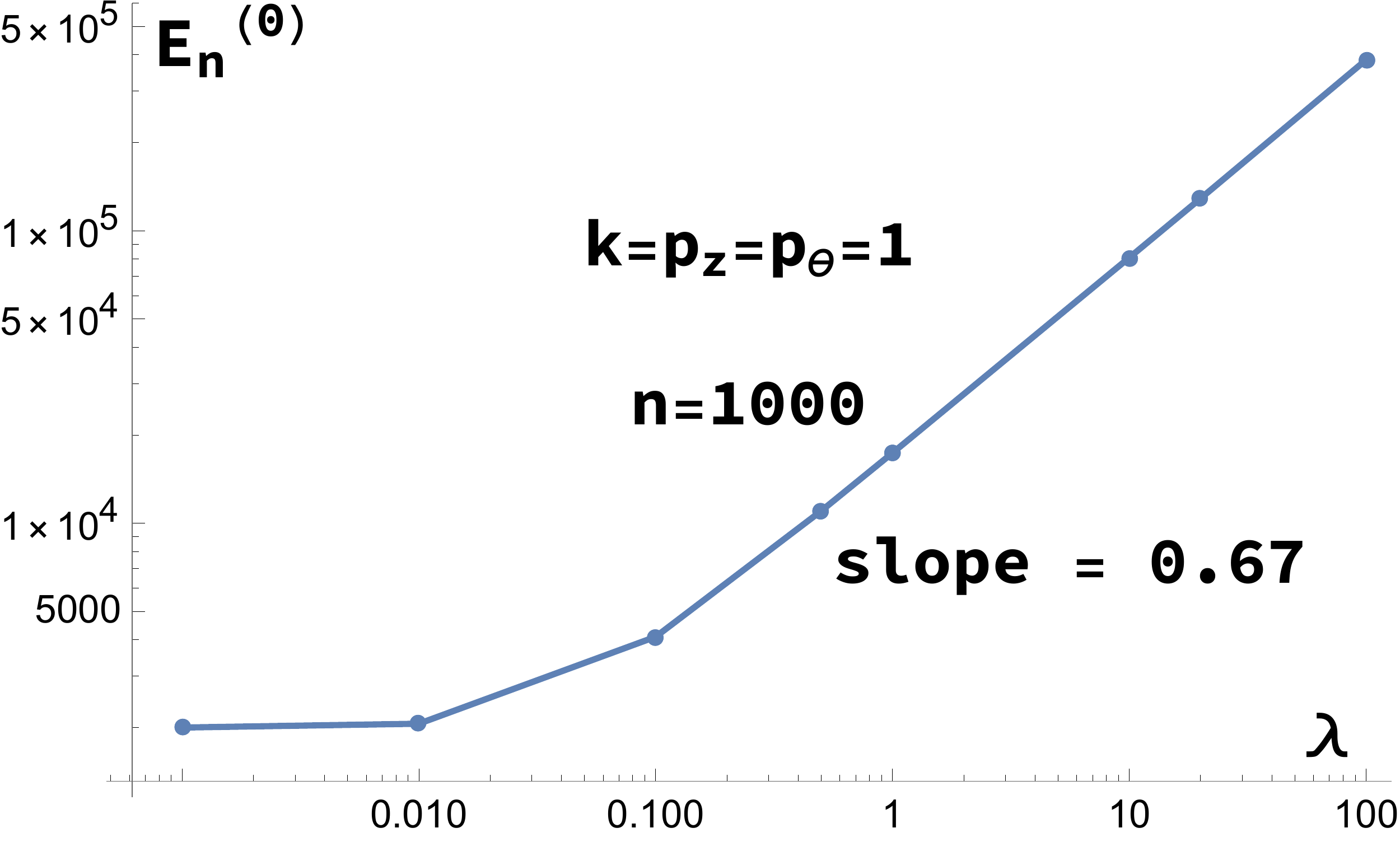}
		\caption{}
		\label{f:EvsLambda-k-1}
		\end{subfigure}
		\qquad 
		\begin{subfigure}[t]{5cm}
		\centering
		\includegraphics[width=6cm]{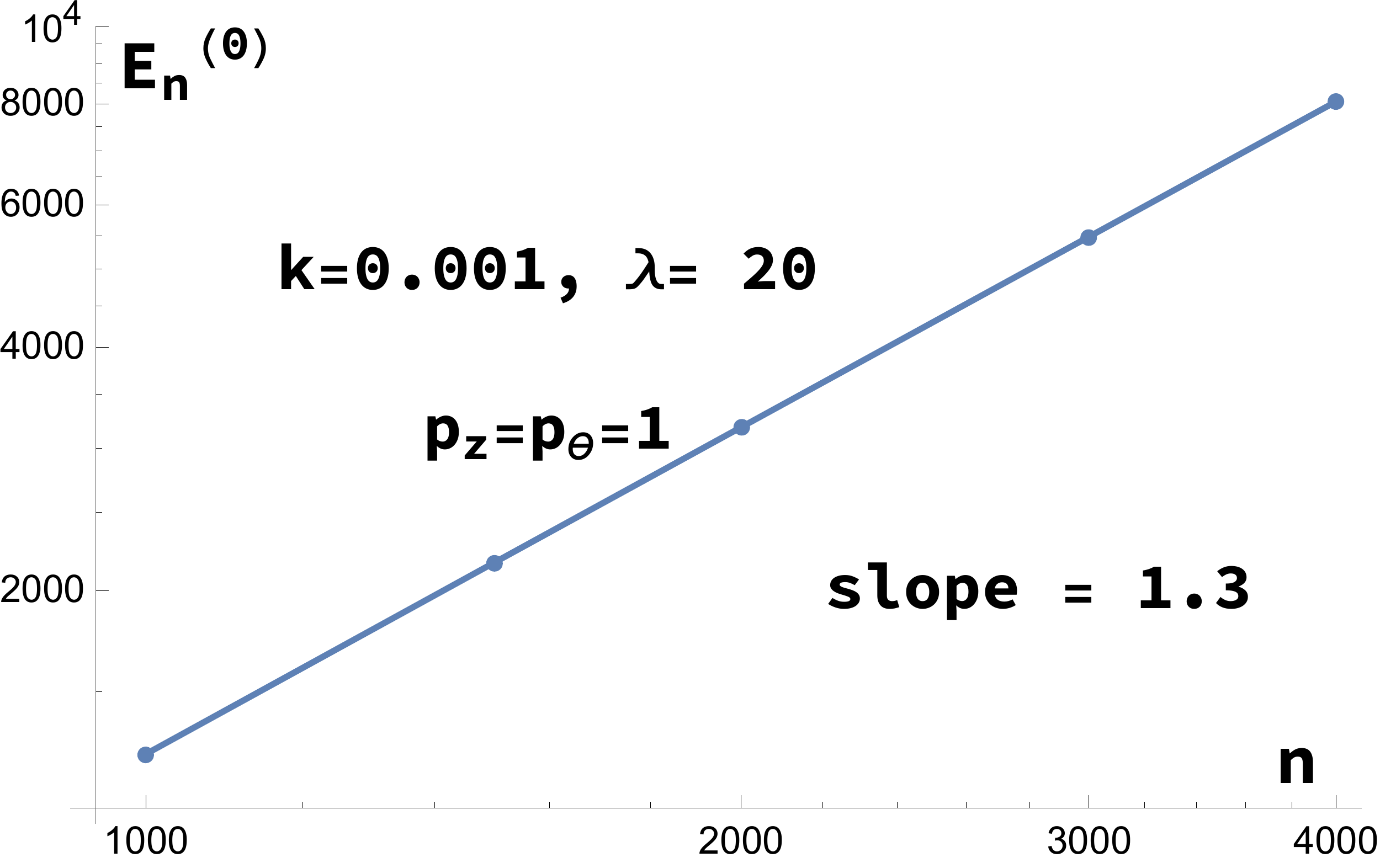}
		\caption{}
		\label{f:EvsN-la-20}
		\end{subfigure}
	\caption{ \footnotesize (a) A $\log$-$\log$ plot of $E_n^{(0)}$ vs $k$, has slope $0.66$ implying the dispersion relation $E_n^{(0)} \propto k^{2/3}$. (b) A $\log$-$\log$ plot of $E_n^{(0)}$ vs $\la$ has an asymptotic slope $0.67$, implying that $E_n^{(0)} \propto \la^{2/3}$. In (a) and (b) the power $2/3$ is independent of the values of $p_z$ and $p_\tht$. (c) $\log E_n^{(0)}$ vs $\log n$ displays a power law $E_n^{(0)} \propto n^{\gamma}$ for large level number $n$.}
	\label{f:EvsK-EvsLambda-and-EvsN}
\end{figure}

\begin{enumerate}

% \item For small $\la$, the energy $E^{(0)}_n$ depends linearly on $k$ ($1\leq k \leq 15$) and $n$ $(100 \leq n \leq 4000)$ in agreement with the weak coupling spectrum of (\ref{e:weak-coupling-WKB-spectrum}). 

\item (a) For moderate to large $\la$ $(1 \leq \la \leq 100)$ and a range of $k$ values $(0.001 \leq k \leq 20)$, $E^{(0)}_n(k) \propto k^{2/3}$ for large $n$ ($900 - 4000$) as shown in Fig.~\ref{f:EvsK-la-1}. Thus, for highly excited states, there is a crossover from $k^1$ to $k^{2/3}$ behaviour as $\la$ increases. 
	
	\vspace{0.25cm}
	
	(b) For fixed $k$, large $n$ $(900 - 4000)$ and $1 \leq \la \leq 100$, $E^{(0)}_n(\la) \propto \la^{2/3}$ as in Fig.~\ref{f:EvsLambda-k-1}.  
	
	\vspace{0.25cm}
	
	(c)  Furthermore, we find that for $1 \leq \la \leq 100$ and a range of $k$ $(0.1 \leq k \leq 20)$, the spectrum of highly excited states $(n \gtrsim 1000)$ depends only on the product $\la k$. In fact, for fixed $n$, $E^{(0)}_n \propto (\la k)^{2/3}$ for $1 \leq \la k \leq 100$. The power $2/3$ is independent of the values of $p_z$ and $p_\tht$. Though they were working with an incorrect radial equation, Rajeev and Ranken \cite{R-R}  proposed this power law based on a dimensional argument that we have not been able to reproduce. The fact that the energies of highly excited states at strong coupling can depend on $\la$ and $k$ only through their product was anticipated in (\ref{e:radial-equation-highly-excited}). It follows that for moderate to large coupling and highly excited states,
	 \beq
	E_n \approx \kappa_n(l) \frac{\hbar^2}{\mu m^2} \tl \sig^{2/3} = \kappa_n(l) \frac{(\la k \hbar^2)^{2/3}}{\mu}.
	\eeq
Numerically, the exponent $2/3$ is found to be insensitive to the values of $p_z$ and $l$, though the proportionality factor $\kappa_n$ could depend on them. However, from (\ref{e:radial-equation-highly-excited}), we infer that $\kappa_n$ must be independent of $p_z$ when $E_n \gg p_z^2/2 \mu$.

\item For moderate to large $\la$ and large $n$, $E_n^{(0)}  \propto n^\gamma$, where $\gamma = \g(k,\la,\pi_z,l)$ (e.g., $\gamma \approx 1.3$ for $\la = 20, k = 0.001, p_z =1$ and $l=1$, see Fig.~\ref{f:EvsN-la-20}).

\end{enumerate}

%-----------------------------------------
\section{Unitary representation of nilpotent algebra}
\label{s:unitary-representation-of-nilpotent-Lie-algebra}
%------------------------------------------

Here, we exploit the quantization of the RR model in the Darboux coordinates $(R_a, k P_a)$ of \S \ref{s:Quantum-RR-model} to obtain a representation of the Poisson algebra of the $L$-$S$ variables (\ref{e:L-and-S-variables}) of the model. Classically, the latter satisfy the step-3 nilpotent Poisson brackets \cite{R-R, G-V-1}
	\beq
	\{ L_a, L_b \} = 0, \quad \{S_a, S_b \} = \la \eps_{abc} L_c \quad \text{and}  \quad \{ S_a, L_b \} = -\eps_{abc} K_c.
	\eeq
Interestingly, as shown in Appendix \ref{E:RR-equations-as-Euler-equations-for-a-nilpotent-Lie-algebra}, the classical EOM of the RR model can be interpreted as Euler equations for this nilpotent Lie algebra.

To relate $L$ and $S$ to the Darboux coordinates, recall from (\ref{e:L-and-S-variables}) that $L = [K,R] + m K$ and $S = \dot R + K/ \la$, where $K = i k \sig_3/2$. The momenta conjugate to $R_a$ are $kP_{1,2} = \dot R_{1,2} \pm (\la m k R_{2,1}/2)$ and $kP_3 = \dot R_3 + \la k (R_1^2 + R_2^2)/2$ (see Eqn.~(\ref{e:conjugate-momenta})). Thus, we may write
	\beqs
	L_1 &=& k R_2, \quad L_2 = -k R_1, \quad L_3 = -m k, \quad K_{1,2} = 0, \quad K_3 = -k \cr 
	S_1 &=& k P_1 - \frac{\la}{2}  k m R_2, \quad S_2 = k P_2 + \frac{\la}{2} k m R_1 \quad \text{and} \quad S_3 = k P_3 - \frac{k}{\la} - \frac{\la k}{2}(R_1^2 + R_2^2). \qquad
	\label{e:L-and-S-component-form-Darboux-coordinates}
	\eeqs
In the quantum theory, we wish to represent $L, S$ and $K$ as Hermitian operators on a Hilbert space obeying the commutation relations obtained by the replacement $\{ A , B \} \to (1/i \hbar) [A, B]$:
	\beq
	\quad [L_a, L_b] = 0, \quad [S_a, S_b] = i \hbar \la \eps_{abc} L_c,  \quad \text{and} \quad [S_a, L_b ] = - i \hbar \eps_{abc} K_c.
	\label{e:commutation-relations}
	\eeq
More explicitly, the nonzero commutation relations among the generators are:
	\beqs
	[L_1, S_2] &=& -i \hbar K_3, \quad [L_2, S_1] = i \hbar K_3, \quad [S_1, S_2] = i \hbar \la L_3, \cr
	[S_1, S_3] &=& -i \hbar \la L_2 \quad \text{and} \quad [ S_2, S_3 ] = i \hbar \la L_1.
	\label{e:nilpotent-Lie-algebra}
	\eeqs
We now use (\ref{e:L-and-S-component-form-Darboux-coordinates}) and the Schr\"odinger representation $R_a = x_a$ and $kP_a = p_a = -i\hbar \pdr_a$ to obtain a representation of this nilpotent Lie algebra:	
	\beqs
	L_1 &=& k y, \quad L_2 = -k x, \quad L_3 = -m k I, \quad K_{1,2} = 0, \quad K_3 = -k I \cr 
	S_1 &=& -i \hbar \pdr_{x} - \frac{\la}{2} m k y, \quad S_2 = - i \hbar \pdr_{y} + \frac{\la}{2} m k x \quad \text{and} \quad
	S_3 = -i \hbar \dd{}{z} - \frac{k}{\la} I- \frac{\la k}{2}(x^2 + y^2), \qquad
	\label{e:representation-of-nilpotent-algebra}
	\eeqs
where $I$ is the identity. These Hermitian operators on the Hilbert space $L^2(\mathbb{R}^3_{xyz})$ give us an infinite-dimensional unitary representation of the nilpotent algebra (\ref{e:commutation-relations}). 

The dynamics of the quantum RR model is specified by a Hermitian and positive Hamiltonian that is a quadratic form on this Lie algebra:
	\beq
	H  = \frac{S_a^2 + L_a^2}{2} + \frac{k S_3}{\la} + \frac{k^2 }{2 \la^2} = -\tr \left[ \left(S - \frac{K}{\la} \right)^2 + L^2 \right].
	\label{e:Hamiltonian-RR-LS-variables-quantum}
	\eeq
Using (\ref{e:commutation-relations}), we find the quadratically nonlinear Heisenberg EOM
	\beq
	\dot S_a = (1/i\hbar) [S_a, H ] =  \la \eps_{abc} S_b L_c  \quad \text{and} \quad %= \la [S, L]_a
	\dot L_a = (1/i \hbar) [L_a, H ] = \eps_{abc} K_b S_c. %= [K, S]_a.
	\label{e:L-S-Heisenberg-EOM}
	\eeq
	
{\fl \bf Reducibility of the representation:} As in the classical theory, $L_3 = -m k$ and $\mathfrak{c} k^2 = (L_1^2 + L_2^2 + L_3^2)/2 + k S_3/\la$ are Casimir operators of the nilpotent commutator algebra (\ref{e:commutation-relations}). We may represent them as differential operators on $L^2(\mathbb{R}^3_{xyz})$:
	\beq
	L_3 = -m k I \quad \text{and} \quad \mathfrak{c} k^2 =  \left(k^2 m^2/2 - k^2/\la^2 \right) I - (i \hbar k/\la) \pdr_{z}.
	\label{e:Casimir-operators}
	\eeq
Both $I$ and $\pdr_z$ commute with all the operators in (\ref{e:representation-of-nilpotent-algebra}), as the latter do not involve $z$. Thus, the representation (\ref{e:representation-of-nilpotent-algebra}) is reducible with invariant subspaces given by the simultaneous eigenspaces of $L_3$ and $\mathfrak{c}$. The latter carry sub-representations labelled by the eigenvalues of $L_3$ and $\mathfrak{c} k^2$. The eigenvalue problem for $\mathfrak{c} k^2$ 
	\beq
	\left[- \frac{i \hbar k}{\la} \dd{}{z} + \left(\frac{m^2 k^2}{2} - \frac{k^2}{\la^2} \right) I \right]  \psi(x,y,z) =  \frac{k p_z}{\la} \psi(x,y,z),
	\eeq
leads to the eigenfunctions $\psi(x, y, z) = F(x,y) \exp(i p_z z/\hbar)$, corresponding to the eigenvalue $k p_z/ \la$. Thus, the representation decomposes as a direct sum of sub-representations labelled by the two real numbers $m$ and $p_z$. Since $F(x,y)$ is an arbitrary function, these sub-representations on $L^2(\mathbb{R}^2_{xy})$ are infinite dimensional, with the generators represented as:
	\beqs
	L_1 &=& k y, \quad L_2 = -k x, \quad L_3 = -m k I, \quad K_3 = -k I \cr 
	S_1 &=& -i \hbar \pdr_x - \frac{\la}{2} m k y, \quad S_2 = - i \hbar \pdr_{y} + \frac{\la}{2} m k x \quad \text{and} \quad
	S_3 = \left( p_z - \frac{k}{\la}\right) I - \frac{\la k}{2}(x^2 + y^2). \qquad
	\label{e:irreducible-representation-of-nilpotent-Lie-algebra}
	\eeqs 
These continue to satisfy the step-3 nilpotent Lie algebra (\ref{e:nilpotent-Lie-algebra}). Since there are no additional Casimirs, (\ref{e:irreducible-representation-of-nilpotent-Lie-algebra}) now furnishes a unitary irreducible representation of  (\ref{e:nilpotent-Lie-algebra}).

%------------------------------
\section{Discussion}
\label{s:Discussion}
%------------------------------

In this paper, we discussed some aspects of the quantum Rajeev-Ranken model for screw-type waves in a 1+1D scalar field theory by interpreting it as a 3D cylindrically symmetric anharmonic oscillator using our Darboux coordinates. This oscillator is unconventional as it has a rotational energy proportional to $L_z$ in addition to a quartic potential. Our approach complements the electromagnetic viewpoint in \cite{R-R}. We quantize the model, find that the quantum theory involves two dimensionless parameters and separate variables in the Schr\"odinger eigenvalue problem. In Ince's classification, the radial equation is of type $[0,1,1_6]$, with one non-elementary regular singular point at $r=0$ and one irregular singular point  of species $6$ (or Poincar\'e rank $3$) at $r = \infty$. It may be obtained from an ODE with $10$ elementary regular singular points via suitable confluences. By contrast, the  Lam\'e equation for ellipsoidal harmonics and the Heun equation arise through confluences of $5$ and $8$ elementary regular singular points. Four-term recurrence relations for coefficients in Frobenius series for energy eigenfunctions are obtained. The hypergeometric, Lam\'e and Heun equations involve 2, 3 and 3-term recurrence relations. The asymptotic behaviour of energy eigenfunctions is found in terms of the Poincar\'e rank of the irregular singularity. The radial equation is examined at weak coupling, strong coupling (where it is argued that $\la$ and $k$ must enter only through their product for highly excited states) and in a double-scaling strong coupling limit, allowing us to get a glimpse of the dispersion relations satisfied by the quantized screwons of the RR model in these limits. While the weak coupling spectrum is determined explicitly, this has not yet been possible for larger values of the coupling $\la$. Moreover, the double-scaling limit  has its limitations, as it selects out only those states whose energies grow quadratically with $\la$. Thus, we also consider the semi-classical approximation, where we express the WKB quantization condition in terms of elliptic integrals. However, we have not yet been able to invert them to find the WKB spectrum, except at weak coupling. Numerical inversion leads us to an interesting fractional power-law dispersion relation $E_n \propto (\la k)^{2/3}$ for highly energetic quantized screwons at moderate and large values of the coupling. The dispersion relations in the double-scaling and WKB strong coupling limits are different; in the former $\hbar \to \infty$ while $\hbar \to 0$ in the latter. In another direction, viewing the EOM of the RR model as Euler equations for a nilpotent linear Poisson algebra, we use  our canonical quantization to find an infinite dimensional irreducible unitary representation of this nilpotent Lie algebra.

%------------------------------
\vspace{0.25cm}
%------------------------------

{\fl \bf Resurgent expansion for WKB spectrum:} It may be possible to obtain the WKB spectrum of the RR model in a semi-classical expansion by extending the results in \cite{Z-J-1, Z-J-2}. In these papers, the authors obtained resurgent expansions with instanton corrections (based on exact quantization conditions) for energy eigenvalues of the quartic double well potential in 1D. They also relate the WKB quantization conditions for the latter (with an additional linear symmetry breaking term) to those of an O(2) anharmonic oscillator with a negative quartic term and the radial equation (see Eq.~(7.4) in \cite{Z-J-2}):
	\beq
	 -\left( \frac{{\rm d^2}}{{\rm d}r^2}  + \frac{1}{r} \frac{{\rm d}}{{\rm d}r} - \frac{l^2}{r^2} \right) \psi(r) + (r^2 - 2 g r^4) \psi(r) =  {\mathcal E} \psi(r).
	 \label{e:zinn-justin-jentschura-radial-eqn}
	\eeq 
Notice that the coefficient of $r^2$ has been normalized to 1 (in \cite{Voros}, the coefficient of $r^2$ was assumed to vanish). By contrast, in our radial equation (\ref{e:radial-eigenvalue-problem-for-quad-quartic}), the RR model coupling $\la$ and wavenumber $k$ appear in both the coefficients $\al $ and $\beta$ (\ref{e:alpha-and-beta-of-quartic-potential}) in the potential $U(r) = \al r^2 + \beta r^4$, with $\beta > 0$. Nevertheless, we may relate (\ref{e:radial-eigenvalue-problem-for-quad-quartic}) to (\ref{e:zinn-justin-jentschura-radial-eqn}) by rescaling $r \to r_1 = \sqrt{\al} r$:
	\beq
	- \frac{\hbar^2 \al}{2 \mu} \left( \frac{{\rm d^2}}{{\rm d}r_1^2}  + \frac{1}{r_1} \frac{{\rm d}}{{\rm d}r_1} - \frac{l^2}{r_1^2} \right) \rho(r_1)+ \left(r_1^2 + \frac{\beta}{\alpha^2} r_1^4 \right) \rho  =  \left( E -\frac{p_z^2}{2 \mu} - \frac{\hbar l \la k m}{2 \mu} - \frac{k^2 m^2}{2}  \right) \rho.
	\label{e:rescaled-radial-equation-RR-model}
	\eeq
For this to match (\ref{e:zinn-justin-jentschura-radial-eqn}), we choose $\mu = \al \hbar^2/2$. Using (\ref{e:alpha-and-beta-of-quartic-potential}), this leads to the quadratic equation 
	\beq
	16 \mu^2 - 4 k^2 \hbar^2 \mu -  \hbar^2(\la^2 k^2 m^2  - 4  \la k p_z)  = 0,
	\eeq
with solutions 
	\beq
	\mu_{\pm} = \frac{1}{8} \left( k^2 \hbar^2 \pm \hbar \sqrt{\hbar^2 k^4 + 4(\la^2 k^2 m^2 - 4 \la k p_z)} \right). 
	\eeq
Restricting to $\al > 0$ ensures that $\mu_{\pm} \in \mathbb{R}$ and $\mu_{+} > 0$ (more generally, the $r_1$-axis would be rotated in the complex plane). If we identify
	\beq
	g(\la, m, k, p_z) = -\frac{\beta}{2 \alpha(\mu)^2} \quad \text{and} \quad  {\mathcal E} = E - \frac{p_z^2}{2 \mu} - \frac{\hbar l \la k m}{2 \mu} - \frac{k^2 m^2}{2},
	\label{e:g-alpha-relation-on-pz}
	\eeq
then we may map solutions of (\ref{e:zinn-justin-jentschura-radial-eqn}) to those of (\ref{e:rescaled-radial-equation-RR-model}), with the reversed sign of $g$ being dealt with through suitable analytic continuation. Here, the coupling $g$ depends on $\la, k, m$ and $p_z$ through $\al$ and $\beta$. Thus, using the methods of \cite{Z-J-1, Z-J-2}, it may be possible to obtain a resurgent expansion for the RR model energy levels $E_n$. We hope to address this in future work.

It is also noteworthy that recently, the authors of \cite{I-M-S} connect the exact WKB periods of certain 1D quantum mechanical systems with polynomial potentials, to a system of thermodynamic Bethe ansatz equations, extending previous work on the ODE/IM correspondence \cite{D-T}. It would be interesting to explore this connection for the potentials that arise in the RR model.	

In another direction, we would like to compare our results on the screw-type waves of the RR model with those for waves that arise from a reduction of the pseudodual principal chiral model. A quantum $r$-matrix formulation for the RR model would also be desirable. The connection between the strong coupling limit of the model and sub-Riemannian geometry (and its quantum counterpart) pointed out in \cite{R-R} is also of interest. Unlike solitons, screwons are not spatially localized. One wonders whether there is an analog of multi-soliton states and their statistics that is applicable to screwons. In this context, it is noteworthy that despite appearances, the trajectory on the $R_1$-$R_2$ plane is quasiperiodic and {\it not} $4 \pi$ periodic, as Fig.1 of \cite{R-R} might suggest. Finally, we hope that a detailed understanding of the quantum RR model (including its spectrum beyond the semi-classical approximation) will help shed light on the dynamics and high energy behavior of the parent nilpotent scalar field theory. 
 
%------------------------------
\vspace{0.25cm}
%------------------------------

{\fl \bf Acknowledgements:} We would like to thank  G. Date, D. Jatkar, A. Laddha, P. Mohile, S. G. Rajeev and P. Ramadevi for useful discussions and references. This work was supported in part by the Infosys Foundation and grants (MTR/2018/000734, CRG/2018/002040) from the Science and Engineering Research Board, Govt. of India.

%-----------------------------
\appendix
\numberwithin{equation}{section}
%-----------------------------

%-----------------------------
\section{Poincar\'e rank and Ince's classification of ODEs}
\label{A:Singularities-of-second-order-ordinary-differential-equations}
%-----------------------------

{\fl \bf Singularities of second order ODEs:} The radial equations (\ref{e:radial-equation-dimensionless-coupling}) and (\ref{e:strong-coupling-radial-equation-dimensionless}) are $2^{\rm{nd}}$ order homogeneous linear ODEs with rational coefficients. To place them in context, we summarize some features of the class of ODEs:
	\beq
	y''(z) + p(z) y'(z) + q(z) y(z) = 0,
	\label{e:second-order-ODE}
	\eeq
where $p$ and $q$ are meromorphic functions on the complex plane. If both $p(z)$ and $q(z)$ are regular at $z_0$, then $z_0$ is an ordinary point; any other point is a singular point. A point $z_0 \neq \infty$ is a regular singularity if at least one of $p$ or $q$ has a pole at $z_0$ such that if $p$ has a pole, it is a simple pole and if $q$ has a pole, it is at most a double pole. On the other hand, $z_0 \neq \infty$ is an irregular singularity if either $p$ has at least a double pole or $q$ has at least a triple pole \cite{A-S-S-W-P-D}.

The nature of $z_0 = \infty$ is determined by putting $\zeta = 1/z$ in (\ref{e:second-order-ODE}):
	\beq
	\frac{d^2 y}{d \zeta^2} + \left[ \frac{2}{\zeta} - \frac{1}{\zeta^2} p \left(\ov{\zeta} \right) \right] \frac{d y}{d \zeta} + \frac{1}{\zeta^4} q \left( \ov {\zeta} \right) y =  0.
	\label{e:transform-linear-ODE}
	\eeq
$z = \infty$ is an ordinary point/regular/irregular singularity of (\ref{e:second-order-ODE}), if $\zeta = 0$ is a corresponding point of (\ref{e:transform-linear-ODE}). In other words, $z = \infty $ is an ordinary point if the Laurent series of $p$ and $q$ around $z = \infty$ are of the form $p(z) = 2/z + \cdots$ and $q(z) = q_4/z^4 + \cdots$. On the other hand, $z = \infty$ is a regular singularity if the Laurent series of $p$ and $q$ around $z = \infty$ satisfy either 
%---------------------------
\begin{enumerate}

\item  $p(z) = p_1/z + \cdots$ and $q(z) = q_2/z^2 + q_3/ z^3 + \cdots$ with $q_2$ and $q_3$ not both zero or

\item $p(z) = p_1/z + \cdots$ with $p_1 \neq 2$ and $q(z) = q_4/z^4 + \cdots$. 

\end{enumerate}

%---------------------------
Finally, $z = \infty$ is an irregular singularity if it is neither an ordinary nor a regular singular point. Alternatively, it is an irregular singularity if either the Laurent series of $p$ around $z = \infty$ contains at least one nonnegative power ($z^0, z^1 \cdots$) or that of $q$ contains at least one power larger than $-2$ ($1/z, z^0, \cdots$). For example, $y'' + a y' + b y = 0$ with constants $a$ and $b$ not both zero has an irregular singularity at $z = \infty$, while every other point is an ordinary point. Indeed, the solution $y = c_1 e^{r_1 z} + c_2 e^{r_2 z}$ has an essential singularity at $z = \infty$. If $a$ and $b$ are both zero, then $y = c_1 z + c_2$ has a simple pole at $z = \infty$ which is a regular singular point. In general, at an ordinary point, the solution of (\ref{e:second-order-ODE}) is analytic. At a regular singular point, it is either analytic, has a pole of finite order or an algebraic or logarithmic branch point singularity. At an irregular singular point, the solution typically has an essential singularity \cite{B-O}. 

At a regular singularity $z_0 \neq \infty$ (if $z_0 = \infty$ we work with $\zeta = 1/z$), we may expand the solution in a Frobenius series $y = (z - z_0)^\rho \sum_{0}^{\infty} y_n (z - z_0)^n$ with the possible exponents $\rho = \rho_{1,2}$ determined by the indicial equation
	\beq
	\rho^2 + (A - 1) \rho + B = 0 \quad \text{where} \quad 
	A = \lim_{z \to z_0} (z - z_0) p(z)  \quad \text{and} \quad B = \lim_{z \to z_0} (z - z_0)^2 q(z).
	\label{e:Indicial-equation-general-form}
	\eeq
In fact, $A = B = 0$ iff $z_0$ is an ordinary point while $z_0$ is a regular singularity iff the limits exist with $A$ and $B$ not both zero. Moreover, if $|\rho_1 - \rho_2| = 1/2$, then $z_0$ is called an elementary regular singular point. Otherwise it is nonelementary \cite{Ince, W-W}. Though they may change the location of $z_0$, fractional linear transformations preserve the nature of $z_0$ (ordinary, elementary/nonelementary regular or irregular). However, other nonlinear transformations such as  $z \to \al z + \beta z^2$  do not, in general, preserve the nature of $z_0$.

%---------------------
\vspace{0.25cm}
%---------------------

{\fl \bf Poincar\'e rank and species:} The Poincar\'e {\it rank} of a singular point is a measure of its irregularity. For definiteness, suppose $z_0 = \infty$ is a singular point, then its rank $g$ is defined as 
	\beq
	g = 1 + {\rm max} \left( K_1, \frac{K_2}{2} \right) \quad \text{where} \quad p(z) = \mathcal{O}(z^{K_1}) \quad \text{and}  \quad q(z) = \mathcal{O}(z^{K_2}) \quad \text{as} \quad z \to \infty.
	\label{e:rank-of-irregular-singularity}
	\eeq
If $z_0 = \infty$ is a regular singularity, its rank is either zero or a negative (half) integer, while for an irregular singularity it can be $1/2, 1, 3/2, \cdots$. Notably, it is possible to double the rank of an irregular singular point via the transformation $z = w^2$ and  restrict to integer ranks. For example, the equation $ y'' + (1/z) y' + (1/z) y = 0$ has a rank 1/2 irregular singularity at $z = \infty$, but becomes $y''- (1/2w) y' + 4 y = 0$, which has a rank 1 irregular singularity at $w = \infty$. Moving away from $\infty$, a singular point $z_0 \neq \infty$ of (\ref{e:second-order-ODE}) is said to have the rank $g = 1+ {\rm max} (K_1, K_2/2)$ if $p(z)$ and $q(z)$ have poles of order $K_1 + 2$ and $K_2 + 4$ respectively (see Eqn.~(\ref{e:transform-linear-ODE})). The {\it species} of an irregular singularity is twice its rank \cite{A-S-S-W-P-D}. 

The rank controls the asymptotic behaviour of solutions to (\ref{e:second-order-ODE}) at an irregular singular point. If $z_0 = \infty$ is an irregular singular point of integer rank $g$, then 
	\beq
	y (z) \sim \exp[A_g z^g + A_{g-1} z^{g-1} + \cdots + A_1 z] Y(z), \quad \text{where} \quad
	Y(z) = z^{-\rho}\sum_{n \geq 0} y_n z^{- n}.
	\label{e:rank-asymptotic-relation-solution}
	\eeq
The rank of an irregular singularity is reminiscent of the genus of an entire function.

%----------------------------------
\vspace{0.25cm}
%----------------------------------

{\fl \bf Invariance of rank:} Though the quadratic transformation $z = w^2$ doubles the rank of an irregular singularity, there is a class of transformations that preserve it. In fact, under a fractional linear transformation $w = (az + b)/(cz + d)$, the coefficients of (\ref{e:second-order-ODE}) remain meromorphic and the rank of a singularity remains unchanged, though its location may be altered. On the other hand, under a linear change of dependent variable $y(z) = F(z) a(z)$, (\ref{e:second-order-ODE}) becomes
	\beq
	a''(z) + \left( 2 \frac{F'(z)}{F(z)} + p(z) \right) a'(z) + \left( \frac{F''(z)}{F(z)} + p(z) \frac{F'(z)}{F(z)} + q(z) \right) a(z) = 0.
	\label{e:linear-transformed-ODE}
	\eeq
To ensure that (\ref{e:linear-transformed-ODE}) has meromorphic coefficients, we will restrict to functions of the form $F = z^{\mu} R_1 e^{R_2}$, where $\mu$ is real and $R_{1,2}(z)$ are rational functions. For definiteness, let us suppose that $z = \infty$ is a rank $g$ irregular singular point of (\ref{e:second-order-ODE}). Furthermore, suppose $R_2(z) \sim z^n$ as $z \to \infty$. Then we find that $z = \infty$ continues to be a rank $g$ irregular singularity of (\ref{e:linear-transformed-ODE}) provided $n \leq g$. In particular, there is no restriction on $\mu$ or $R_1$. This restriction on $n$ is expected from the connection between the rank and the asymptotic behaviour in (\ref{e:rank-asymptotic-relation-solution}).

%-----------------------------
\vspace{0.25cm}
%-----------------------------

{\fl \bf Confluences of elementary regular singularities:} It is possible to create irregular singular points through the confluence of regular ones  \cite{Ince}. For instance, the coalescence of 2 elementary regular singular points produces a nonelementary one, while the merger of 3 elementaries gives an irregular singularity of species 1 (rank 1/2). More generally, an irregular singularity of species $r$ is formed by the coalescence of $r+2$ elementary regular singularities. This motivates the classification of regular singularities into elementary and nonelementary \cite{Ince, W-W}. 

%-----------------------------
\vspace{0.25cm}
%-----------------------------

{\fl \bf Ince's classification:} Ince classifies ODEs of type (\ref{e:second-order-ODE}) based on the number and nature of singularities. Eqn.~(\ref{e:second-order-ODE}) is said to be of type $[a, b, c_i, d_j, \cdots]$, if $a$ is the number of elementary regular singular points, $b$ is the number of nonelementary regular singular points, and $c_i, d_j, \cdots$ are the number of irregular singularities of species $i, j, \cdots$. For example, the hypergeometric equation is denoted $[0, 3, 0]$: it has 3 nonelementary regular singularities at $z = 0, 1$ and  $\infty$ and can be obtained from an equation of type $[6,0,0]$ through suitable confluences. The confluent hypergeometric equation is denoted $[0, 1, 1_2]$. It has  a regular (nonelementary) singularity at zero and an irregular singularity of rank 1 at $\infty$ formed by the coalescence of regular singularities at $1$ and $\infty$. The Heun equation ($[0, 4, 0]$) has 4 nonelementary regular singular points \cite{A-S-S-W-P-D}. When two of them coalesce, we get the confluent Heun equation $([0, 2, 1_2])$ with an irregular singularity of rank 1. The biconfluent Heun equation $([0, 1, 1_4])$ has an irregular singularity of rank 2 formed by the merger of 3 nonelmentary regular singular points. The Lam\'e equation for ellipsoidal harmonics is of type $[3, 1, 0]$ and it has 3 elementary regular singularities and 1 nonelementary regular singularity at infinity. Finally, our radial equation (\ref{e:radial-equation-dimensionless-coupling}) (and its strong coupling limits (\ref{e:radial-equation-highly-excited}) and  (\ref{e:strong-coupling-radial-equation-dimensionless})) is of type $[0,1,1_6]$, since $K_1 = -1$ and $K_2 = 4$ in (\ref{e:rank-of-irregular-singularity}) and is a confluent form of an equation with 10 elementary regular singular points. The ellipsoidal Lam\'e and Heun equations arise from confluences of 5 and 8 elementary regular singular points. As a matter of terminology, we note that Whittaker and Watson \cite{W-W} refer to an equation with 5 elementary regular singularities as a generalized Lam\'e equation, while Ince \cite{Ince} calls an equation with any number of elementary regular singular points a generalized Lam\'e equation.

%----------------------------
\section{Asymptotic behaviour of the radial wavefunction}
\label{B:Asymptotic-behaviour-of-the-radial-wavefunction}
%----------------------------

The radial Schr\"odinger equation (\ref{e:radial-equation-dimensionless-coupling}) of  the quantum RR model in dimensionless variables
	\beq
	 \rho''(\tl r) + \ov{\tl r} \rho'(\tl r) - \left( \frac{l^2}{\tl r^2}  + \frac{\tl \al}{\tl \hbar^2} \tl r^2 + \frac{\tl \beta}{\tl \hbar^2} \tl r^4 + \frac{l \tl \la}{\tl \hbar} - \frac{\tl E_1}{\tl \hbar^2}  \right) \rho(\tl r) = 0
	 \label{e:radial-equation-appendix}
	\eeq
has an irregular singular point at $\tl r = \infty$. To find the asymptotic form of $\rho(\tl r)$, we put $\rho(\tl r) = \exp(S(\tl r))$ anticipating exponential behaviour. In terms of $S(\tl r)$, (\ref{e:radial-equation-appendix}) becomes:
	\beq
	S''(\tl r) + (S'(\tl r))^2 +  \ov{\tl r} S'(\tl r) - \left( \frac{l^2}{\tl r^2}  + \frac{\tl \al}{\tl \hbar^2} \tl r^2 + \frac{\tl \beta}{\tl \hbar^2} \tl r^4 + \frac{l \tl \la}{\tl \hbar} - \frac{\tl E_1}{\tl \hbar^2}  \right) = 0.
	\label{e:radial-equation-dimensionless-change-of-variable}
	\eeq
Now we make the `slowly varying' assumption $|S''(\tl r)| \ll (S'(\tl r))^2$ as $\tl r \to \infty$, which will be seen to be self-consistent. For large $\tl r$, the quartic term in (\ref{e:radial-equation-dimensionless-change-of-variable}) dominates, so the `asymptotic radial equation' is $S'(\tl r)^2  \sim  \tl \beta \tl r^4/\tl \hbar^2$. This implies 
	\beq
	S'(\tl r) \sim \pm \left(\sqrt{\tl \beta} /\tl \hbar \right) \, \tl r^2  \quad \text{or } \quad S(\tl r) = \pm \left(\sqrt{\tl \beta} /3 \tl \hbar \right) \, \tl r^3 + c(\tl r),
	\label{e:S-from-asymptotic-radial-equation}
	\eeq
where the constant of integration $c$ is allowed depend on $\tl r$, in order to allow for subleading behaviour as $\tl r \to \infty$. For consistency, we must have $|c(\tl r)| \ll \sqrt{\tl \beta} \tl r^3/ 3 \tl \hbar$ as $\tl r \to \infty$. For normalizability, $\rho(\tl r) \to 0$ as $\tl r \to \infty$. Thus, we must choose the negative sign for $S(\tl r)$ in (\ref{e:S-from-asymptotic-radial-equation}). Substituting this in (\ref{e:radial-equation-dimensionless-change-of-variable}) we get
	\beq
	c''(\tl r) + c'(\tl r)^2 - \left( \frac{2 \sqrt{\tl \beta}}{\tl \hbar} \tl r^2 - \ov{\tl r} \right) c'(\tl r) - \frac{3 \sqrt{\tl \beta}}{\tl \hbar}  \tl r - \frac{l^2}{\tl r^2} - \frac{\tl \al}{\tl \hbar^2} \tl r^2 - \frac{l \tl \la}{\tl \hbar} + \frac{\tl E_1}{\tl \hbar^2}  = 0.
	\eeq
As before, as $\tl r \to \infty$ we may use the inequalities $|c''(\tl r)| \ll 2 \sqrt{\tl \beta}  \tl r / \tl \hbar$ and $c'(\tl r)^2 \ll \sqrt{\tl \beta} \tl r^2 |c'(r)|/ \tl \hbar$ to obtain an asymptotic equation for $c(\tl r)$:
	\beqs
	\left(2 \sqrt{\tl \beta}/\tl \hbar \right) \tl r^2 c'(\tl r) &\sim& -\left(3 \sqrt{\tl \beta}/\tl \hbar\right) \tl r - (\tl \al/\tl \hbar^2) \tl r^2,\cr
	 \text{which implies} \quad c(\tl r) &\sim& -(3/2) \ln r - \left(\tl \al/2 \sqrt{\tl \beta} \tl \hbar \right)\tl r + {\rm constant}.
	\eeqs
This gives the leading asymptotic behaviour of $\rho(\tl r)$:
	\beq
	\rho(\tl r) \sim \exp\left( -\frac{\sqrt{\tl \beta}}{\tl \hbar} \left( \frac{\tl r^3}{3} + \frac{ \tl \alpha \tl r}{2 \tl \beta} \right) \right) \tl r^{-3/2} a(\tl r), \quad \text{where} \quad a(\tl r) \sim \mathcal{O}(1) \quad \text{as} \quad \tl r \to \infty.
	\label{e:asymptotic-behaviour-rho-generic-coupling}
	\eeq
In the double-scaling strong coupling limit of  \S \ref{s:Weak-and-strong-coupling-limits-of-the-Schrodinger-eigenvalue-problem}, $\tl \al/ \tl \beta \to 1$ and $\sqrt{\tl \beta}/\tl \hbar  \to \tl g/2$. If we view (\ref{e:asymptotic-behaviour-rho-generic-coupling}) as a change of dependent variable, the resulting equation for $a(\tl r)$ has an irregular singularity at $\tl r = \infty$, just as (\ref{e:radial-equation-appendix}) did.

%----------------------------
\section{Frobenius method for double-scaling strong coupling limit}
\label{C:Frobenius-method-for-strong-coupling-limit-Local-analysis}
%----------------------------

Here, we consider a series expansion for the radial wavefunction $\rho(\tl r)$ around the regular singular point $\tl r = 0$ of the radial equation  (\ref{e:strong-coupling-radial-equation-dimensionless}), $\rho(\tl r)  = \sum_{n = 0}^{\infty} \rho_n \tl r^{\eta + n}$. Substituting this in (\ref{e:strong-coupling-radial-equation-dimensionless}) gives
	\beqs
	\sum_{n=0}^\infty \rho_n (\eta + n)(\eta + n -1) \tl r^{\eta + n -2} &+& \frac{1}{\tl r}  \sum_{n=0}^\infty \rho_n ( \eta + n ) \tl r^{\eta + n - 1} \cr
	&-& \left( \frac{l^2}{\tl r^2} + \frac{\tl g^2}{4} (\tl r^2 + \tl r^4) - \tl g^2 \tl E_2 + l \tl g \right) \sum_{n = 0}^\infty \rho_n \tl r^{\eta + n} = 0. \quad
	\eeqs
We rewrite this equation as 
	\beq
	\left[ \sum_{n=0}^\infty ((\eta + n)^2 - l^2) \rho_n  - \frac{\tl g^2}{4}\left( \sum_{n =4}^\infty \rho_{n-4}  + \sum_{n = 6}^\infty \rho_{n-6}  \right)
		+ (\tl g^2 \tl E_2 - l \tl g) \sum_{n = 2}^\infty \rho_{n -2}\right] \tl r^{\eta + n - 2}  = 0.
	\eeq
From this, we get the indicial exponents $\eta = \pm l$. Choosing $\eta = l$ in order that the normalizability condition (\ref{e:normalizability-condition-radial-bound-state}) is satisfied, we get the four-term recurrence relation (\ref{e:recurrence-relations-around-zero-rho}).

%-----------------------------
\section{Elliptic integral for WKB quantization condition}
\label{D:Computation-of-elliptic-integral}
%-----------------------------

The LHS of the WKB quantization condition (\ref{e:quantization-condition-cubic}) can be expressed as a sum of complete elliptic integrals of the first three kinds: $K, E$ and $\Pi$. Multiplying and dividing the integrand by the radical in the denominator, we get 
	\beq
	\int_{x_{\rm min}}^{x_{\rm max}} \frac{dx}{2x} \, \frac{(2 \mu E^{(0)} - p_z^2 - p_{\tht} \la m k - k^2 m^2 \mu) x - 2 \mu( \al x^2 + \beta x^3)  - p_{\tht}^2}{ \left[(2 \mu E^{(0)} - p_z^2 - p_{\tht} \la m k - k^2 m^2 \mu) x - 2 \mu( \al x^2 + \beta x^3)  - p_{\tht}^2 \right]^{1/2}} = n \pi \hbar.
	\label{e:quantization-condition-cubic-modified}
	\eeq 
Let the roots of the cubic $-2 \mu \beta x^3 + ...$ be $c < b = x_{\rm min} < a= x_{\rm max} $ with $0 < b,a$. Then using 3.131(5), 3.132(4) and 3.137(5) of \cite{G-R}, we have
	\beqs
	\int_{b}^a \frac{dx}{\sqrt{(a-x)(x-b)(x-c)}} &=& \frac{2}{\sqrt{a-c}} K(\zeta),  \cr
	\int_{b}^a \frac{x \, dx}{\sqrt{(a-x)(x-b)(x-c)}} &=& \frac{2}{\sqrt{a-c}} \left[ (b-c) \Pi(\zeta^2 , \zeta) + c K(\zeta) \right],   \cr
	\int_{b}^a \frac{x^2 \, dx}{\sqrt{(a-x)(x-b)(x-c)}} &=& \frac{4\sqrt{a-c}(a+b+c)}{3} E(\zeta) + \frac{2(a(c - b) + c(b+ 2c))}{3 \sqrt{a-c}} K(\zeta), \cr
	\int_{b}^a \frac{dx}{x \sqrt{(a-x)(x-b)(x-c)}} &=& \frac{2}{c b\sqrt{a-c}} \left[(c-b) \Pi\left( \zeta^2\frac{c}{b}, \zeta \right) + b K(\zeta) \right].
	\eeqs
Here,  $\zeta = \sqrt{(a-b)/(a-c)}$. The energy eigenvalue $E^{(0)}$ enters through the roots $a,b,c$ but we have not been able to invert (\ref{e:quantization-condition-cubic-modified}) to obtain the spectrum $E^{(0)}_n$ explicitly.

%-----------------------------
\section{RR equations as Euler equations for a nilpotent Lie algebra}
\label{E:RR-equations-as-Euler-equations-for-a-nilpotent-Lie-algebra}
%-----------------------------

The EOM of the RR model $\dot L = [K, S]$ and $\dot S = \la [ S, L]$ (see (\ref{e:L-S-Heisenberg-EOM}) and \cite{R-R, G-V-1}) may be viewed as Euler equations for a nilpotent Lie algebra. Indeed, they follow from the quadratic Hamiltonian 
	\beq
	H = -\tr [(S-K/\la)^2 + L^2],
	\eeq
and the step-3 nilpotent Poisson algebra $\mathfrak{n}_3$ (here, $1\leq a,b,c \leq 3$):
	\small
	\beq
	\{ L_a, L_b \} = \{ K_a, K_b \} =  \{ K_a, L_b \} = \{ K_a, S_b \} = 0, \quad \{S_a, S_b \} = \la \eps_{abc} L_c \quad \text{and} \quad \{ S_a, L_b \} = -\eps_{abc} K_c. \quad 	
	\eeq
	\normalsize
This algebra is a central extension by the generators $K_a$ of the step-2 nilpotent algebra
	\beq
	\mathfrak{n}_2: \quad \{ L_a, L_b \} = 0, \quad \{S_a, S_b \} = \la \eps_{abc} L_c, \quad \{ S_a, L_b \} = 0.
	\eeq
The $L_a$ span an abelian ideal $\mathfrak{l}$ of $\mathfrak{n}_2$ with the 3D abelian quotient $\mathfrak{n}_2/\mathfrak{l}$ generated by the $S_a$.  As before, we take $K_3 = -k, K_{1,2} = 0$ so that $\mathfrak{n}_3$ is seven-dimensional with generators $(L_a, S_a)$ and the identity $I$. The Hamiltonian is a quadratic form on this Lie algebra. If we use the basis $L_a, \tl S_a = S_a - K_a/ \la$ and $I$ then the Hamiltonian is 
	\beq
	H = -\tr(\tl S^2 + L^2)
	\eeq
and corresponds to the inverse inertia matrix ${\cal I}_{ij}^{-1} = {\rm diag }(1,1,1,1,1,1, 0)$. The zero eigenvalue of ${\cal I}_{ij}^{-1}$ in the central direction can be made nonzero by adding a constant term to $H$. Thus, the RR model can be viewed as an Euler top for the nilpotent Lie algebra $\mathfrak{n}_3$. Similarly, the RR equations can also be viewed as Euler equations for a  centrally extended Euclidean algebra as mentioned in \cite{G-V-2}; without the central extension, one gets the Kirchhoff model. 

%----------------------------------------

%--------------------

%-------------------
\end{document}